\documentclass[a4paper,11pt]{article}
\pdfoutput=1 % if your are submitting a pdflatex (i.e. if you have
\usepackage{jheppub} % for details on the use of the package, please
\usepackage[T1]{fontenc} % if needed
\usepackage{amssymb,amsbsy,bm,amsmath,psfrag,dsfont}
\usepackage{hyperref}
\usepackage{graphicx}
\usepackage{epsfig}
\usepackage{subfigure}

\makeatletter
\def\@fpheader{\relax}
\makeatother

\newcommand{\beq}[1]{\begin{equation}\label{#1}}
\newcommand{\eeq}{\end{equation}}
\newcommand{\ba}{\begin{array}}
\newcommand{\ea}{\end{array}}

\newcommand{\be}{\begin{equation}}
\newcommand{\ee}{\end{equation}}
\newcommand{\bea}{\begin{eqnarray}}
\newcommand{\eea}{\end{eqnarray}}

\title{\boldmath Determination of Nonlinear Genetic Architecture using Compressed Sensing}

\author[a]{Chiu Man Ho,}
\author[a]{Stephen~D.~H.~Hsu}

\affiliation[a]{Department of
  Physics and Astronomy, Michigan State University,\\
  East Lansing, MI 48824, USA}

\emailAdd{cmho@msu.edu}
\emailAdd{hsu@msu.edu}

\abstract{
\\ \\
\textbf{Background:}
\\
One of the fundamental problems of modern genomics is to extract the genetic architecture of a complex trait from a data set of individual genotypes and trait values. This problem is complicated by the large number of candidate genes, the potentially large number of causal loci, and the likely presence of some nonlinear interactions between different genes. Compressed Sensing methods obtain solutions to under-constrained systems of linear equations. These methods can be applied to the problem of determining the best model relating genotype to phenotype, and generally deliver better performance than simply regressing the phenotype against each genetic variant, one at a time. We introduce a Compressed Sensing method that can reconstruct nonlinear genetic models (i.e., including epistasis, or gene-gene interactions) from phenotype-genotype (GWAS) data. Our method uses L1-penalized regression applied to nonlinear functions of the sensing matrix.
\\ \\
\textbf{Findings:}
\\
The computational and data resource requirements for our method are similar to those necessary for reconstruction of linear genetic models (or identification of gene-trait associations), assuming a condition of {\it generalized sparsity}, which limits the total number of gene-gene interactions. An example of a sparse nonlinear model is one in which a typical locus interacts with several or even many others, but only a small subset of {\it all possible} interactions exist. It seems plausible that most genetic architectures fall in this category.  We give theoretical arguments suggesting that the method is nearly optimal in performance, and demonstrate its effectiveness on broad classes of nonlinear genetic models using both real and simulated human genomes. A phase transition in the behavior of the algorithm indicates when sufficient data is available for its successful application.
\\ \\
\textbf{Conclusion:}
\\
Our results indicate that predictive models for many complex traits, including a variety of human disease susceptibilities (e.g., with additive heritability $h^2 \sim 0.5$), can be extracted from data sets comprised of $n_\star \sim 100 s$ individuals, where $s$ is the number of distinct causal variants influencing the trait. For example, given a trait controlled by $\sim 10$k loci, roughly a million individuals would be sufficient for application of the method.}
\begin{document}
\maketitle
\flushbottom

\section{Background}

Realistic models relating phenotype to genotype exhibit nonlinearity (epistasis), allowing distinct regions of DNA to interact with one another. For example, one allele can influence the effect of another, altering its magnitude or sign, even silencing the second allele entirely.  For some traits, the largest component of genetic variance is linear (additive) \cite{linear}, but even in this case nonlinear interactions accounting for some smaller component of variance are expected to be present. To obtain the best possible model for prediction of phenotype from genotype, or to obtain the best possible understanding of the genetic architecture, requires the ability to extract information concerning nonlinearity from phenotype--genotype (e.g., GWAS) data. In this paper we describe a computational method for this purpose.

Our method makes use of compressive sensing (CS) \cite{CS1,CS2,CS3,CStext}, a framework originally developed for recovering sparse signals acquired from a linear sensor. The application of CS to genomic prediction (using linear models) and GWAS has been described in an earlier paper by one of the authors \cite{GCS}. Before describing the new application, we first summarize results from \cite{GCS}.

Compressed sensing allows efficient solution of underdetermined linear systems:
\begin{equation}
\label{ax}
y = Ax + \epsilon ~~,
\end{equation}
($\epsilon$ is a noise term) using a form of penalized regression. L1 penalization, or LASSO, involves minimization of an objective function over candidate vectors $\hat{x}$:
\begin{equation}
\label{O}
O = \vert \vert y - A \hat{x} \vert \vert_{L2} + \lambda \vert \vert \hat{x} \vert \vert_{L1}~~,
\end{equation}
where the penalization parameter is determined by the noise variance (see results section for more detail). Because $O$ is a convex function it is easy to minimize. Recent theorems \cite{CS1,CS2,CS3,CStext} provide performance guarantees, and show that the $\hat{x}$ that minimizes $O$ is overwhelmingly likely to be the sparsest solution to (\ref{ax}). In the context of genomics, $y$ is the phenotype, $A$ is a matrix of genotypes (in subsequent notation we will refer to it as $g$), $x$ a vector of effect sizes, and the noise is due to nonlinear gene-gene interactions and the effect of the environment.

Let $p$ be the number of variables (i.e., dimensionality of $x$, or number of genetic loci), $s$ the sparsity (number of variables or loci with nonzero effect on the phenotype; i.e., nonzero entries in $x$) and $n$ the number of measurements of the phenotype (i.e., dimensionality of $y$ or the number of individuals in the sample). Then  $A$  is an  $n \times p$  dimensional matrix. Traditional statistical thinking suggests that $n > p$  is required to fully reconstruct the solution  $x$  (i.e., reconstruct the effect sizes of each of the loci). But recent theorems in compressed sensing \cite{CS1,CS2,CS3,CStext,GCS} show that  $n > C s \log p$ (for constant $C$ defined over a class of matrices $A$) is sufficient if the matrix $A$ has the right properties (is a good compressed sensor). These theorems guarantee that the performance of a compressed sensor is nearly optimal -- within an overall constant of what is possible if an oracle were to reveal in advance which $s$  loci out of $p$ have nonzero effect. In fact, one expects a phase transition in the behavior of the method as $n$ crosses a critical threshold $n_{\star}$ given by the inequality. In the good phase ($n > n_{\star}$), full recovery of $x$ is possible.

In \cite{GCS}, it is shown that

a. Matrices of human SNP genotypes are good compressed sensors and are in the universality class of (i.e., have the same phase diagram as) random (Gaussian) matrices. The phase diagram is a function of sparsity $s$ and sample size $n$ rescaled by dimensionality $p$. Given this result, simulations can be used to {\it predict} the sample size threshold for future genomic analyses.

b. In applications with real data the phase transition can be detected from the behavior of the algorithm as the amount of data $n$ is varied. (For example, in the low noise case the mean p-value of selected, or nonzero, components of $x$ exhibits a sharp jump at $n_{\star}$.) A priori knowledge of $s$ is not required; in fact one deduces the value of $s$ this way.

c.  For heritability $h^2 = 0.5$ and $p \sim 10^6$ SNPs, the value of $C  \log  p \sim 30$. For example, a trait which is controlled by $s = 10$k loci would require a sample size of $n \sim 300$k individuals to determine the (linear) genetic architecture (i.e., to determine the full support, or subspace of nonzero effects, of $x$).

Our algorithm for dealing with nonlinear models is described in more detail in the Methods section below. It exploits the fact that although a genetic model $G(g)$ with epistasis depends nonlinearly on $g$, it only depends linearly on the interaction parameters $X$ which specify the interaction coefficients (i.e., $z$ and $Z$ in Eq. (\ref{model})). Briefly, the method proceeds in two Steps:

{\bf Step 1}. Run CS on $(y,g)$ data, using linear model (\ref{yax}). Determine support of $x$: subset defined by $s$ loci of nonzero effect.

{\bf Step 2}. Compute $G(g)$ over this subspace. Run CS on $y = G(g) \cdot X$ model to extract nonzero components of $X$. These can be translated back into the linear and nonlinear effects of the original model (i.e., nonzero components of $z$ and $Z$).
\bigskip

In the following section we show that in many cases Steps 1 and 2 lead to very good reconstruction of the original model (\ref{model}) given enough data $n$. A number of related issues are discussed:

a. When can nonlinear effects hide causal loci from linear regression (Step 1)? In cases of this sort the locus in question would not be discovered by GWAS using linear methods.

b. Both matrices $g$ and $G(g)$ seem to be well-conditioned CS matrices. The expected phase transitions in algorithm performance are observed for both Steps.

c. For a given partition of variance between linear (L), nonlinear (NL) and IID error $\epsilon$, how much data $n_{\star}$ is required before complete selection of causal variants occurs (i.e., crossing of the phase boundary for algorithm performance)? Typically, if Step 1 is successful then with the same amount of data Step 2 will also succeed.

\section{Relation to earlier work}

Here we give a brief discussion of earlier work, with the goal of clarifying what is new and distinctive about our technique.
Reviews of earlier methods aimed at detecting gene-gene interactions can be found in \cite{review1,review2,review3,review4}.

Theoretical results concerning LASSO can be found in, e.g., \cite{theoretical1,theoretical2}, as well as in the well-known work by Candes, Tao, Donoho and collaborators \cite{CS1,CS2,CS3,CStext}. The main point, as discussed in \cite{GCS}, is that despite the beautiful results in this literature there is no theorem that can be specifically applied to matrices formed of genomes, or of nonlinear functions of genomes, because of nontrivial structure: correlations between the genomes of individuals in a population. This structure implies that the matrices of interest deviate at least slightly from random matrices, and have only empirically-determined properties. Therefore, all of the rigorous theorems {\it are merely guides to our intuition or expectations}: empirical results from simulations are necessary to proceed further. As noted in \cite{GCS}, the best one can do is to test the phase transition properties of real genomic matrices and determine whether they are in the same category (``universality class'') as random matrices. This being the case (as verified in \cite{GCS}), one can then use the phase diagram to {\it predict} how much data $n$ is required to determine the optimal genomic model for a given sparsity $s$, dimensionality $p$, and composition of variance (L, NL, IID error).

The key result is that once the phase boundary is crossed to the favorable part of the phase diagram (i.e., sufficient data is available), the support of the candidate vector $\hat{x}$ will coincide with the support of the optimal $x$: i.e., the solution of equation (\ref{O}) in the limit of infinite data, or the sparsest solution to (\ref{ax}). Note that in the presence of noise the specific values of the components of $\hat{x}$ will differ from those of $x$; it is the {\it locations} of the nonzero components (support of $x$) that can be immediately extracted once the data threshold $n_\star$ is surpassed.

\begin{figure}[h!]
\begin{center}
\includegraphics[height=9cm, width=7cm]{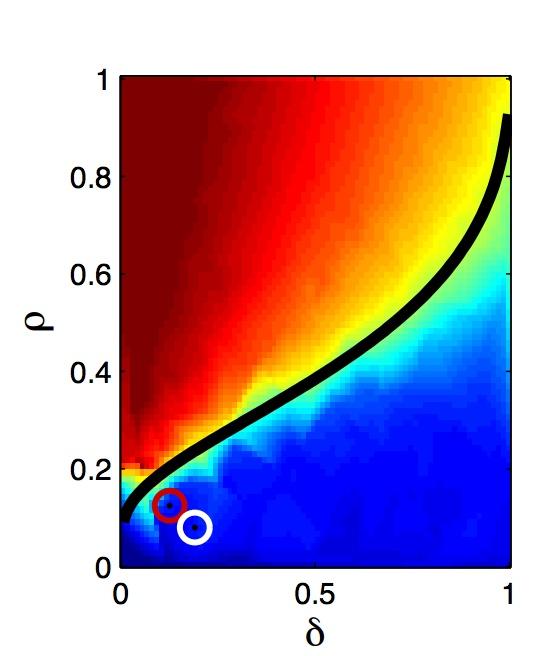}
\caption{Phase diagram found in \cite{GCS} for matrices of human SNP genotypes as a function of $\rho = s/n$ and $\delta = n/p$. This is identical to the diagram found by Dohono and Tanner for Gaussian random matrices in \cite{DT}.}
\label{fig:phase}
\end{center}
\end{figure}

The phase transition properties of compressed sensing algorithms are under-appreciated, especially so given their practical utility.  We refer the reader to pioneering work by Donoho and collaborators \cite{DT,D,DT1,DT2,DT3}. (See also discussion in \cite{CStext} in end notes of Chapter 9.) Briefly, they showed that the phase behavior of CS algorithms is related to high dimensional combinatorial geometry, and that a broad class of random matrices (matrices randomly selected from a specific ensemble, such as Gaussians) {\it exhibit the same phase diagram} (i.e., are in the same universality class) as a function of sparsity and data size rescaled by dimensionality: $(\rho = s/n \, , \, \delta = n/p)$. In \cite{GCS} the same phase boundary was found for matrices of human SNP genotypes, verifying that genomic matrices fall into the same universality class as found in \cite{DT}. In other words, the diagram in  Fig. (\ref{fig:phase}), constructed in \cite{GCS} from simulations with real human genotypes, coincides exactly with the one found in \cite{DT} for broad classes of random matrices.

This paper extends the analysis of \cite{GCS} to nonlinear genomics models incorporating epistasis. We make use of the phase transition to first identify the subspace of loci with nonzero effects and then use L1-penalized regression to fit a nonlinear model defined on this subspace. We find that the nonlinear genomic matrix $G(g)$ is also a good compressed sensor -- it exhibits phase transition behavior similar to that of $g$. In practical terms, our methods should allow the recapture of a large part of nonlinear variance in phenotype prediction.

Some alternative methods for dealing with epistasis are given in \cite{alternative-nonlinear,SLee,TEAM,BOOST,Devlin,Wu} (for an overview of several methods, see \cite{epistasis-review}). The closest proposal to the one examined in this paper is that of Devlin and collaborators \cite{Devlin,Wu}, which uses LASSO and a linear model to first narrow the subspace of interest to lower dimensionality, and then uses an interacting model to estimate nonlinear interactions. The main difference with our work is that (i) \cite{Devlin,Wu} do not exploit the phase transition properties that allow one to determine a critical data threshold beyond which nearly ideal selection of loci has occurred, and (ii) \cite{Devlin,Wu} do not situate their results in the theoretical framework of compressed sensing. In practical terms, as discussed below, we use signals such as the median p-value of candidate loci to determine when the calculation is in the favorable part of the phase diagram.

The motivation of earlier work \cite{alternative-nonlinear,SLee,TEAM,BOOST,Devlin,Wu} seems to be mainly exploratory -- to find specific cases of nonlinear interactions, but not to ``solve'' genetic architectures involving hundreds or thousands of loci. This is understandable given that only a few years ago the largest data sets available were much smaller than what is available today (e.g., hundreds of thousands of individuals, rapidly approaching a million). Our perspective is quite different: we want to establish a lower bound on the total amount of genetic variation (linear plus nonlinear) that can be recaptured for predictive purposes, given data sets that have crossed the phase boundary into the region of good recovery. As we discuss below, our main result is that for large classes of nonlinear genetic architectures one can recapture nearly all of the linear variance and half or more of the nonlinear variance using an amount of data that is somewhat beyond the phase transition for the linear case. We do not claim to have shown that our method performs better than all possible alternative methods. However, given what is known from theoretical analyses in CS, we expect our method to be close to optimal, up to logarithmic corrections and a possible improvement of the constant factor $C$ appearing in the relation $n \sim C s \log p$. For more detailed comparisons of earlier methods, see \cite{epistasis-review}.

\section{Results}

Most of our simulations are performed using synthetic genomes with the minor allele frequency (MAF) restricted to values between 0.05 and 0.5. The synthetic genomes are determined as follows: generate a random population-level MAF $\in (0.05, 0.5)$ for each locus, then populate each individual genome with 0,1,2 SNP values according to the MAF for each locus. Results obtained using synthetic genomes are similar to those obtained from real SNP genomes, as we discuss below.

We study the performance of our algorithm on two specific classes of ``biologically inspired'' models, although we believe our results are generic for any nonlinear models with similar levels of sparsity, nonlinear variance, etc. The models described below are used to generate phenotype data $y$ for a given set of genotypes. Our algorithm then tries to recover the model parameters. Each of the models below is a specific instance of the general class of nonlinear models in equation (\ref{model}); this class is the set of all possible models including linear and bi-linear (gene-gene interaction) effects. The model parameters $\alpha, \beta, \gamma$ used below can be rewritten in terms of the variables $z, Z$ in (\ref{model}).

The first category of models is the block-diagonal (BD) interaction model:
\bea
\label{BD}
y^a = \sum_{i=1}^{s} \alpha_i\,\, g^a_i ~+~ \sum_{i=1}^{s} \beta_i \,\,(\,g^a_i\,)^2 ~+~ \sum_{i=1}^{s-1} \gamma_i \,\,g^a_i\,\, g^a_{i+1} + \epsilon^a ~~~.
\eea
The BD models have $s$ causal loci, each of which has (randomly determined) linear and quadratic effects on the phenotype, as well as mixed terms coupling one locus to another. In biological terms, this model describes a system in which each locus interacts with others in the same block (including itself), but not with loci outside the block.

The second category of models is the ``promiscuous'' (PS) interaction model:
\bea
\label{PS}
y^a = \sum_{i=1}^{s} \alpha'_i\,\, g^a_i  ~+~  \sum_{i=1}^{s'} \beta'_{i}\,\, (\,g^a_{s+i}\,)^2 ~+~ \sum_{i=1}^{s'/2} \gamma'_i \,\,g^a_i\,\, g^a_{s+i} + \epsilon^a ~~~.
\eea
The model has $s$ loci which have linear but no quadratic effect on the phenotype, and $s'$ loci have quadratic but no linear effect on the phenotype. $s'/2$ of the latter type interact with counterparts of the former type. In biological terms, this model has subsets of loci which are entirely linear in effect, some which are entirely nonlinear, and interactions between these subsets.

It is worth remarking that since no large systems of interacting loci are currently understood in real biological systems, one cannot be sure that any specific model (in particular, (\ref{BD}) or (\ref{PS}) above) is realistic. They are merely a way to generate datasets on which to test our method. What we do believe, based on many simulations, is that our method works well on models in which the generalized sparsity (roughly, total number of interactions) is sufficiently small (i.e., far from the maximal case in which all loci interact with all others). In the maximal limit the dimensionality of the parameter space of the resulting model is so large that simple considerations imply that it is intractable. We certainly hope that Nature does not realize such models, and indeed they are implausible as they require essentially every gene to interact nontrivially with every other.

In both models, we fix $\textrm{var}(\epsilon) =0.3$, so the total genetic variance accounted for (i.e., the broad sense heritability) is $0.7$, which is in the realistic range for highly heritable complex traits such as height or cognitive ability (see, e.g., \cite{Hsu,GCTA1,GCTA2}). The total genetic variance can be divided into linear and nonlinear parts; the precise breakdown is determined by the specific parameter values in the model. We chose the probability distributions determining the coefficients so that typically half or somewhat less of the genetic variance is due to nonlinear effects (see figures 2-8). For the BD model, $\alpha_i$, $\beta_i$ and $\gamma_i$ are drawn from normal distributions. Their means are of order unity and positive but not all the same. The standard deviation of $\alpha_i$ is larger than those of $\beta_i$ and $\gamma_i$, but all of them are smaller than their respective means. In particular, we take their means to be $\mu(\alpha) =1.5$, $\mu(\beta)=1.0$ and $\mu(\gamma)=0.5$, and their standard deviations to be $\sigma(\alpha) =0.5$, $\sigma(\beta)=0.2$ and $\sigma(\gamma)=0.1$. We study the cases $s=5, 50, 100$ with $p=10000, 25000, 40000$ respectively. For the PS model, $\alpha'_i$ are drawn from $\{-1, 0, 1\}$. $\beta'_i$ and $\gamma'_i$ are randomly chosen from normal distributions, and are typically of order unity. (In the results shown, negative values of  $\beta'_i$ and $\gamma'_i$ are excluded, but similar results are obtained if negative are also allowed.) We study the cases $\{s=3, s'=2 \}, \{s=30, s'=20 \}, \{s=60, s'=40 \}$ with $p=10000, 20000, 30000$ respectively.

For each of the models, we first perform Step 1 (i.e., run CS on the $(y,g)$ data) with a value of $n$ that is very close to $p$. For such a large sample size Step 1 of the algorithm always works with high precision, and determines a best-fit linear approximation (hyperplane) to the data. From this Step we can calculate the variance accounted for by linear effects, and the remaining variance which is nonlinear. Parameters deduced in this manner are effectively properties of the model itself and not of the algorithm. We denote by $x^\ast$ the resulting $\hat{x}$; this is an ``asymptotic'' linear effects vector that would result from having a very large amount of sample data.

The nonlinear variance is defined as
\bea
\label{NL_variance}
\sigma^2_{\textrm{NL}} \equiv \textrm{var} (y - g \,x^\ast - \epsilon).
\eea
We use this quantity to estimate the penalization $\lambda$ for LASSO: we set $\lambda = \sigma^2_{\textrm{NL}} + \textrm{var}(\epsilon)$. In a realistic case one could set $\lambda$ using an estimate of the additive heritability of the phenotype in question (e.g., obtained via twin or adoption study, or GCTA \cite{GCTA1,GCTA2}). This penalization may be larger than necessary; if so, the required data threshold for the phase transition might be reduced from what we observe.

\begin{figure}[h!]
\begin{center}
\includegraphics[height=7cm, width=9cm]{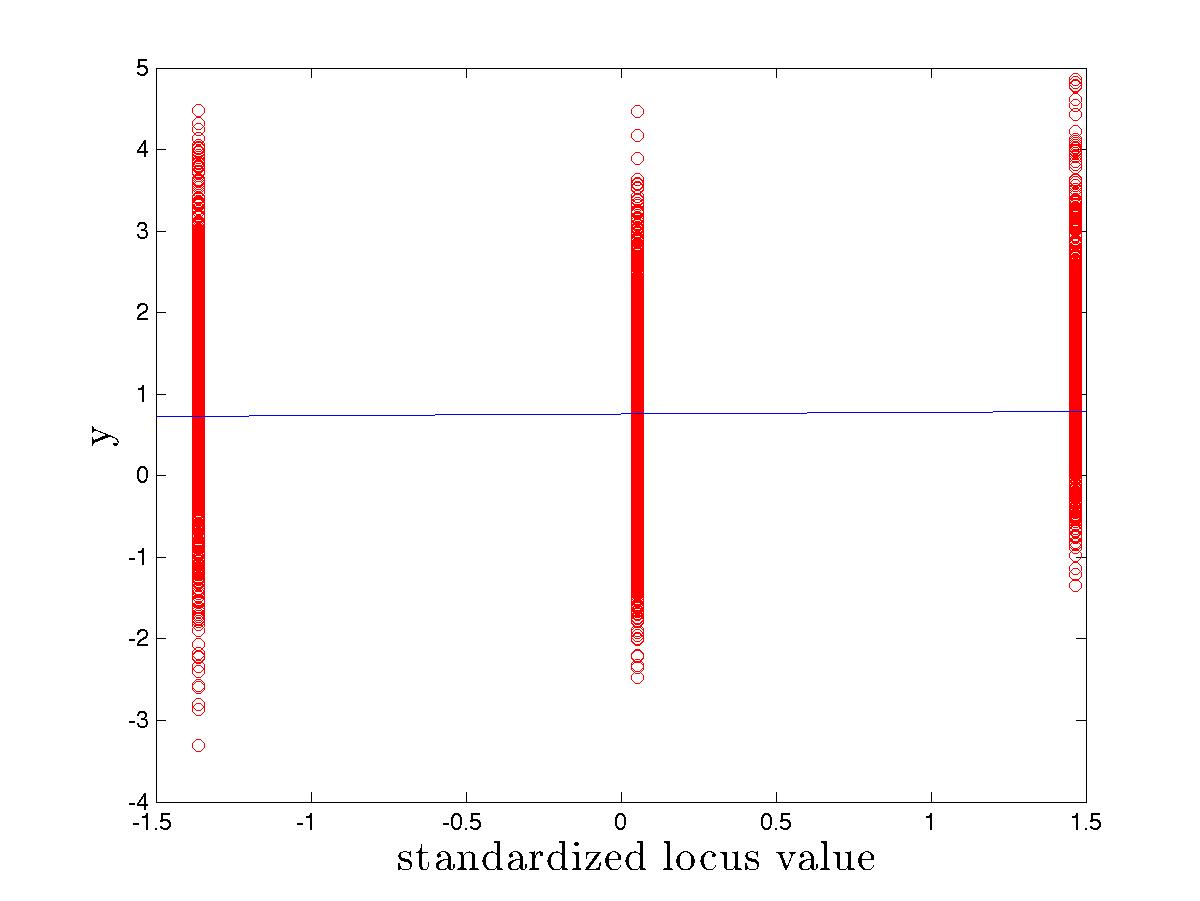}
\caption{Phenotype as a function of standardized locus value. The linear regression (blue line) of phenotype versus this locus value has slope close to zero. PS model with $s+s'=5$.}
\label{fig:locus}
\end{center}
\end{figure}

From $x^\ast$ we know which effects are detectable by linear CS under ideal (large sample size) conditions. In some cases, the nonlinear effects can hide a locus from detection even though at the model level (e.g., in Eqs. (\ref{BD}, \ref{PS})) it has a direct linear effect on the phenotype $y$. This happens if the best linear fit of $y$ as a function of the locus in question has slope nearly zero (see Fig. (\ref{fig:locus})). We refer to the fraction of causal loci for which this occurs as the {\it fraction of model zeros}. These loci are not recoverable from either linear regression or linear CS even with large amounts of data ($n \approx p$). When this fraction is nonzero, the subspace of causal variants that is detected in Step 1 of our algorithm will differ from the actual subspace (in fact, Step 1 recovery of the causal subspace will sometimes fall short of this ideal limit, as realistic sample sizes $n$ may be much less than $p$; see upper right and lower left panels in Fig. (\ref{fig:BD}) and Fig. (\ref{fig:PS})). This is the main cause of imperfect reconstruction of the full nonlinear model, as we discuss below.

\begin{figure}[h!]
\begin{center}
\includegraphics[height=7cm, width=9cm]{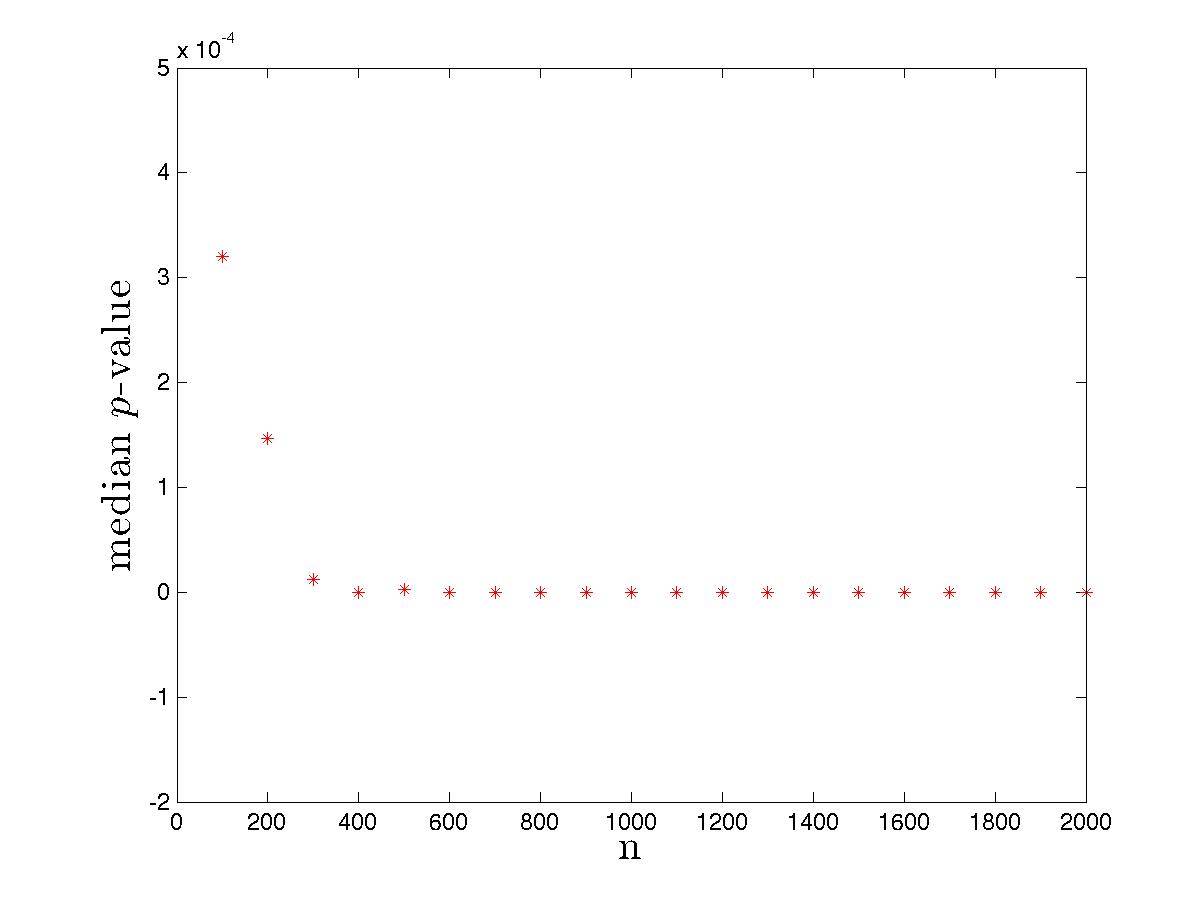}
\caption{The phase transition in median $p$-value as a function of sample size $n$. PS model with $s+s'=5$.}
\label{fig:Pvalue}
\end{center}
\end{figure}

\begin{figure}[t!]
\includegraphics[height=6.0cm, width=8.0cm]{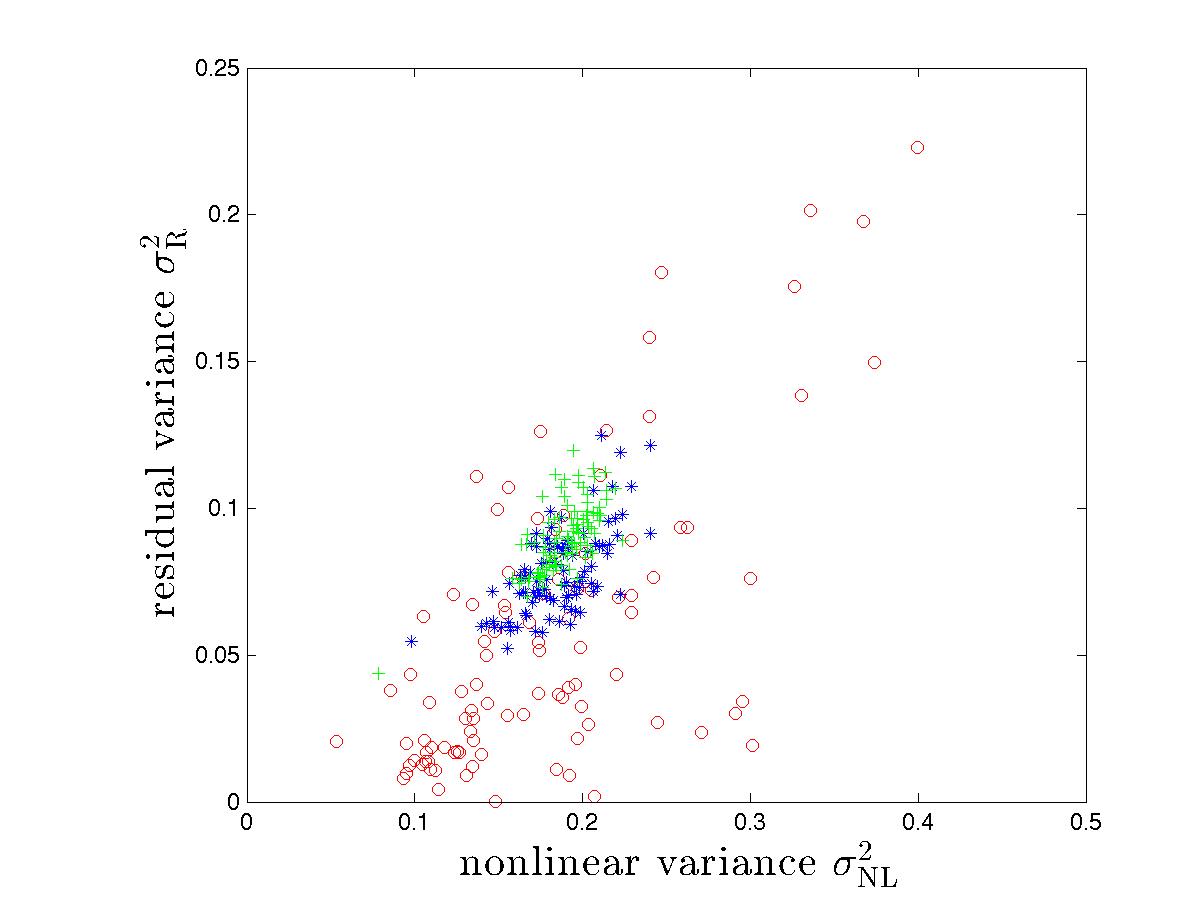}
~~
\includegraphics[height=6.0cm, width=8.0cm]{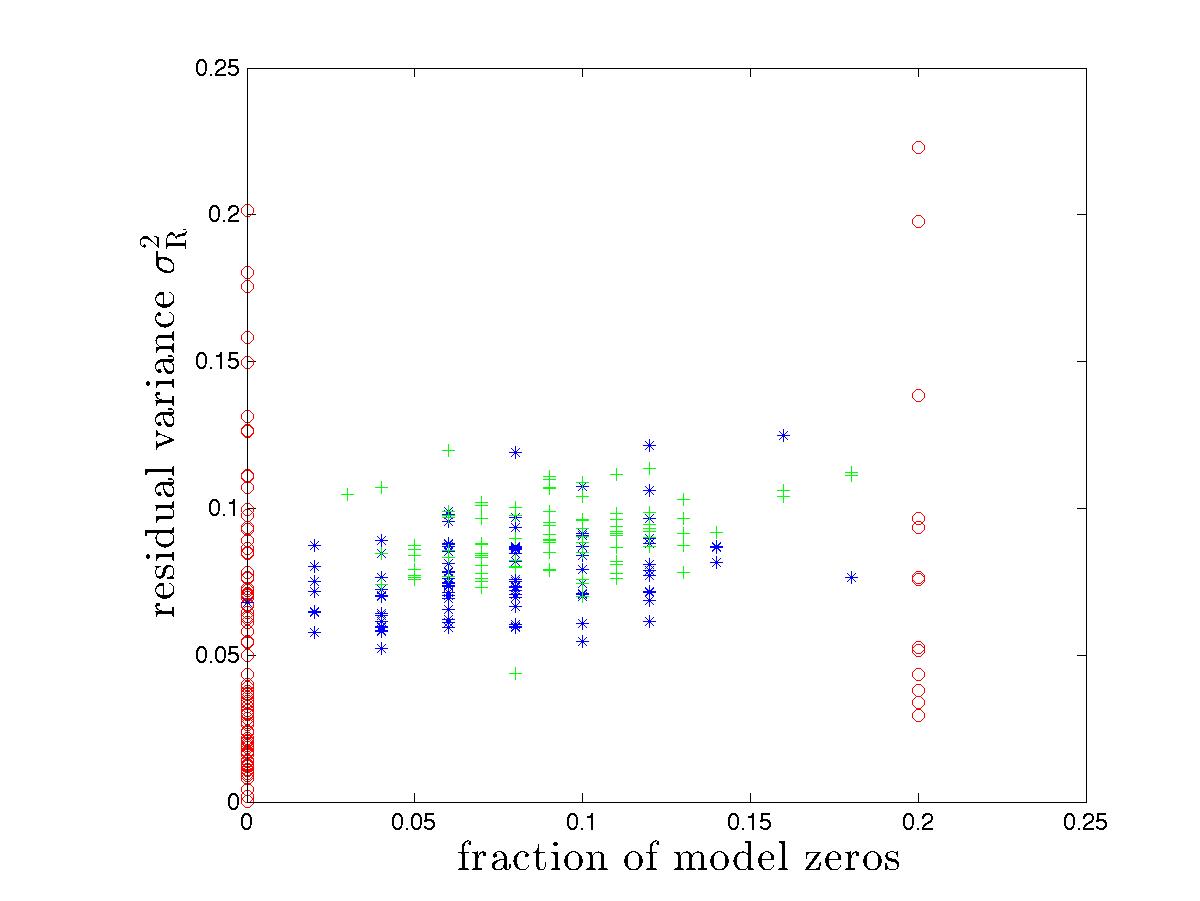}
~~
\includegraphics[height=6.0cm, width=8.0cm]{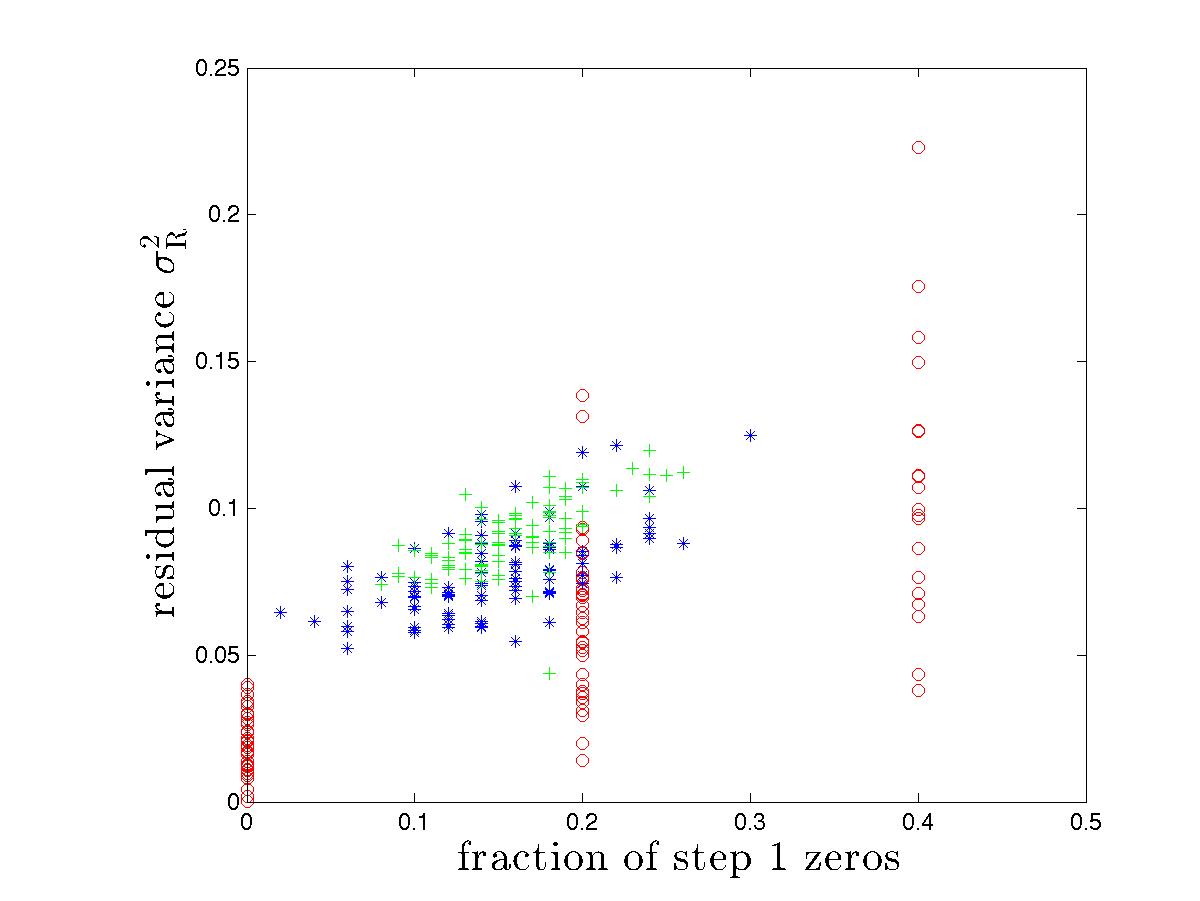}
~~
\includegraphics[height=6.0cm, width=8.0cm]{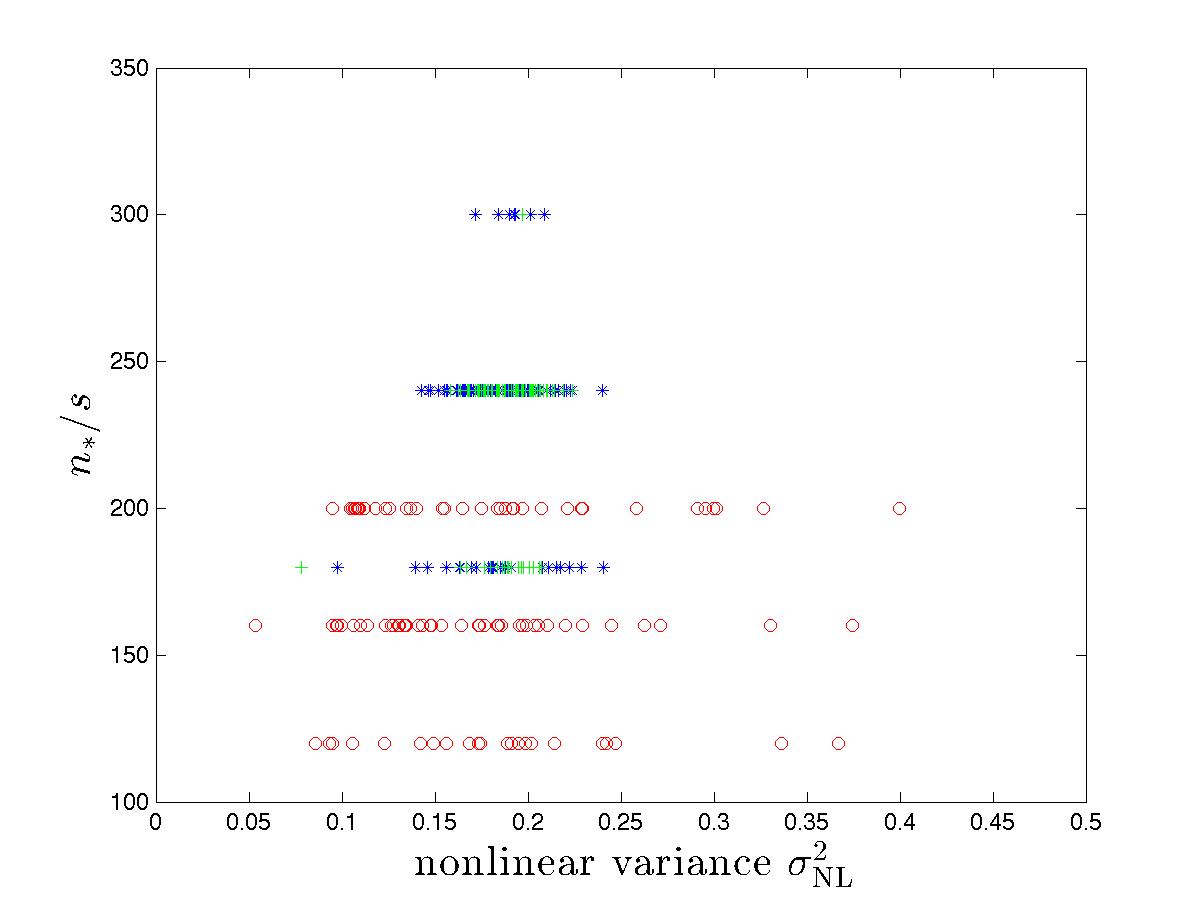}
\caption{BD model with synthetic genomes. Red, blue and green symbols correspond to cases with $s=5, 50, 100$ respectively. Results for 100 runs (i.e., 100 different realizations of the model) are shown for each case.}
\label{fig:BD}
\end{figure}

\begin{figure}[t!]
\includegraphics[height=6.0cm, width=8.0cm]{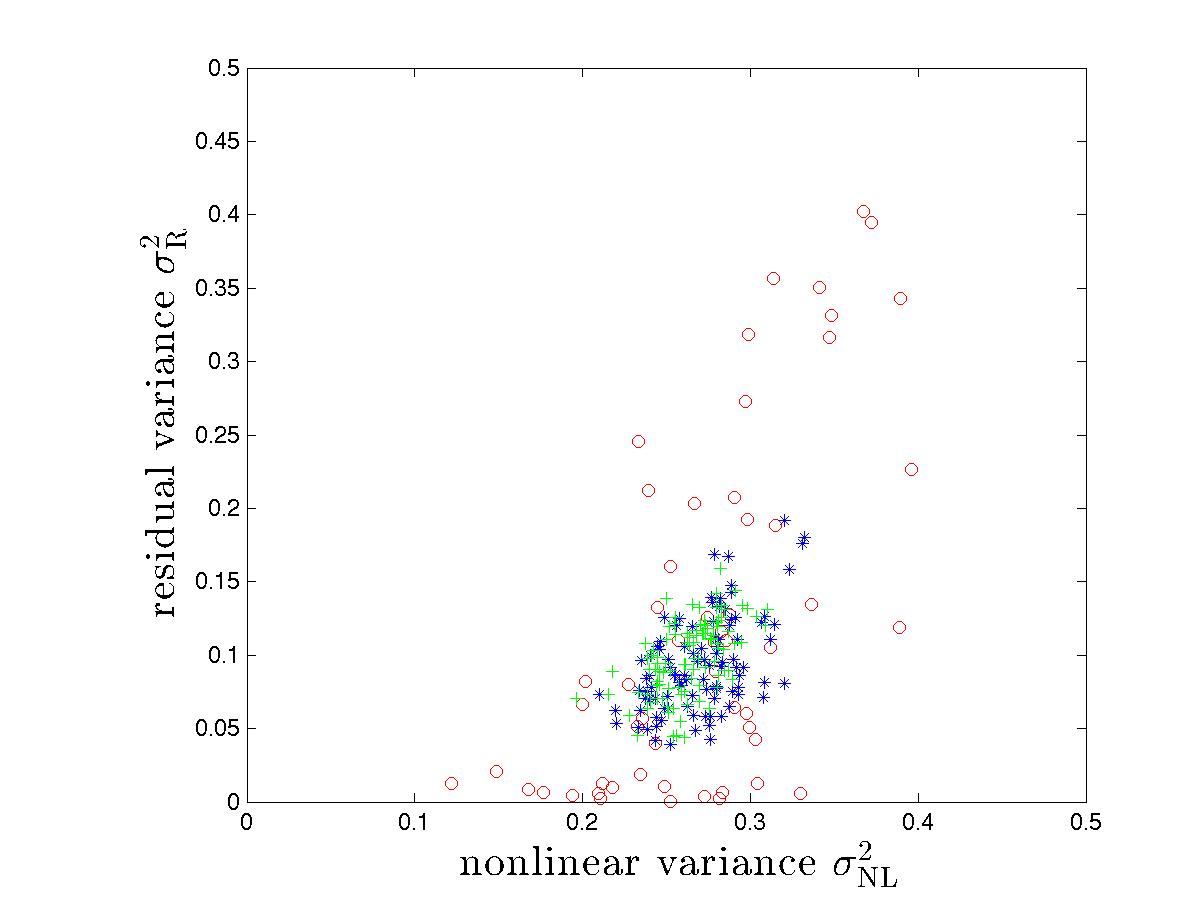}
~~
\includegraphics[height=6.0cm, width=8.0cm]{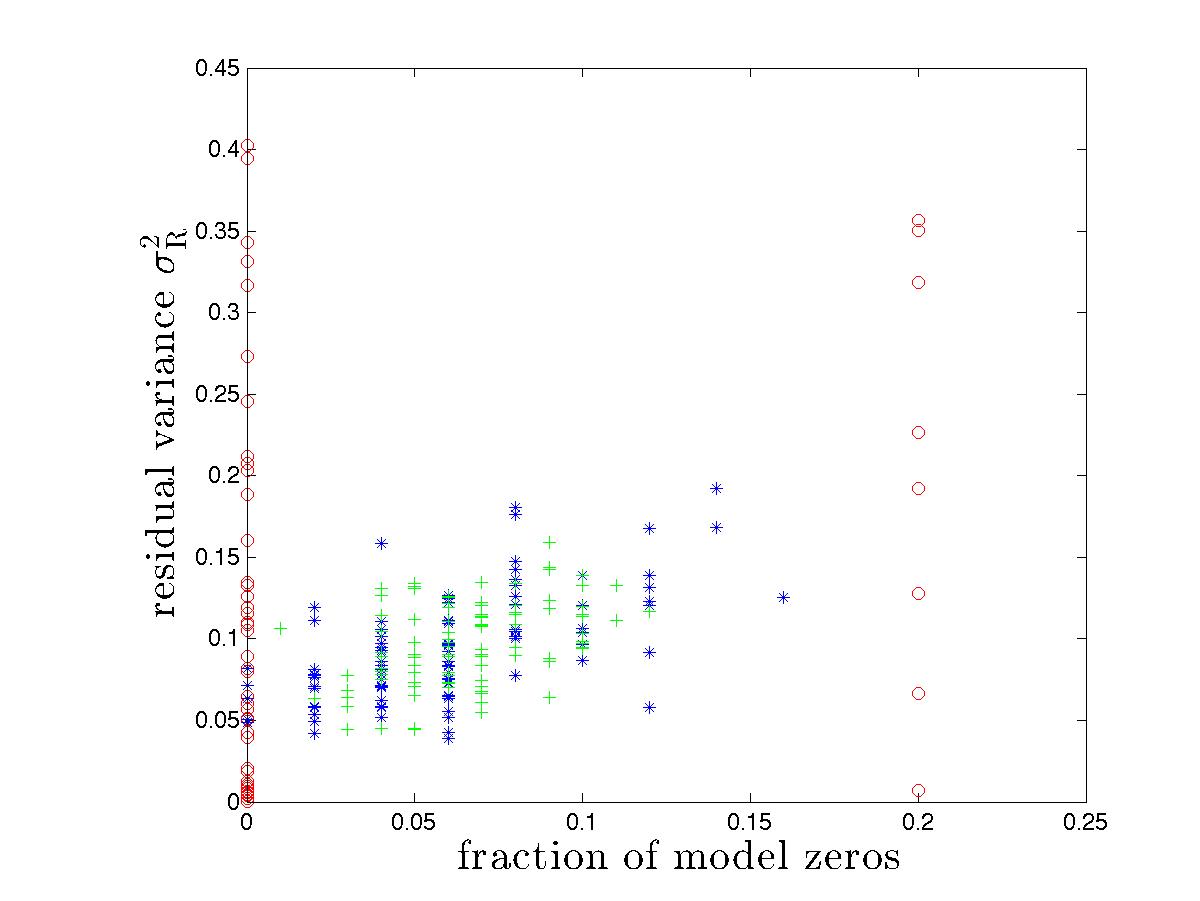}
~~
\includegraphics[height=6.0cm, width=8.0cm]{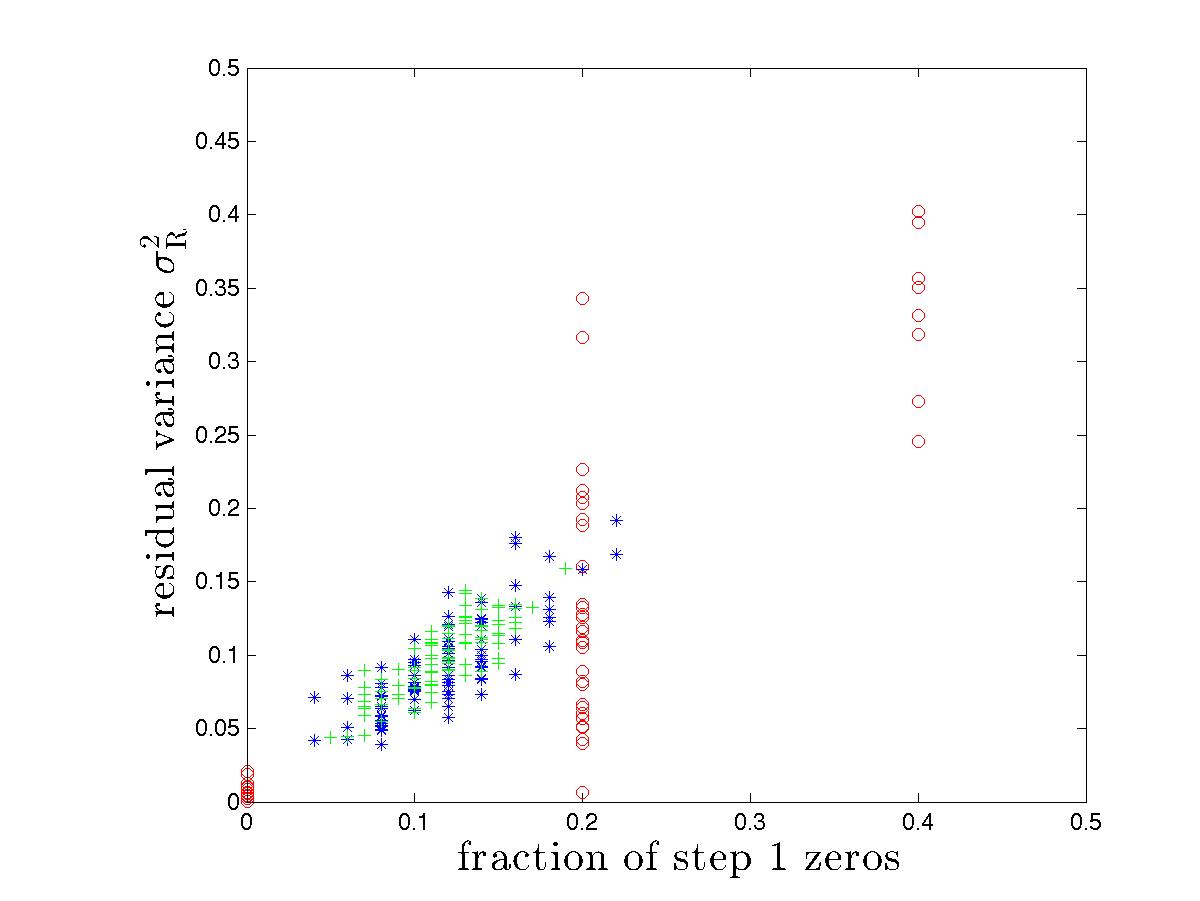}
~~
\includegraphics[height=6.0cm, width=8.0cm]{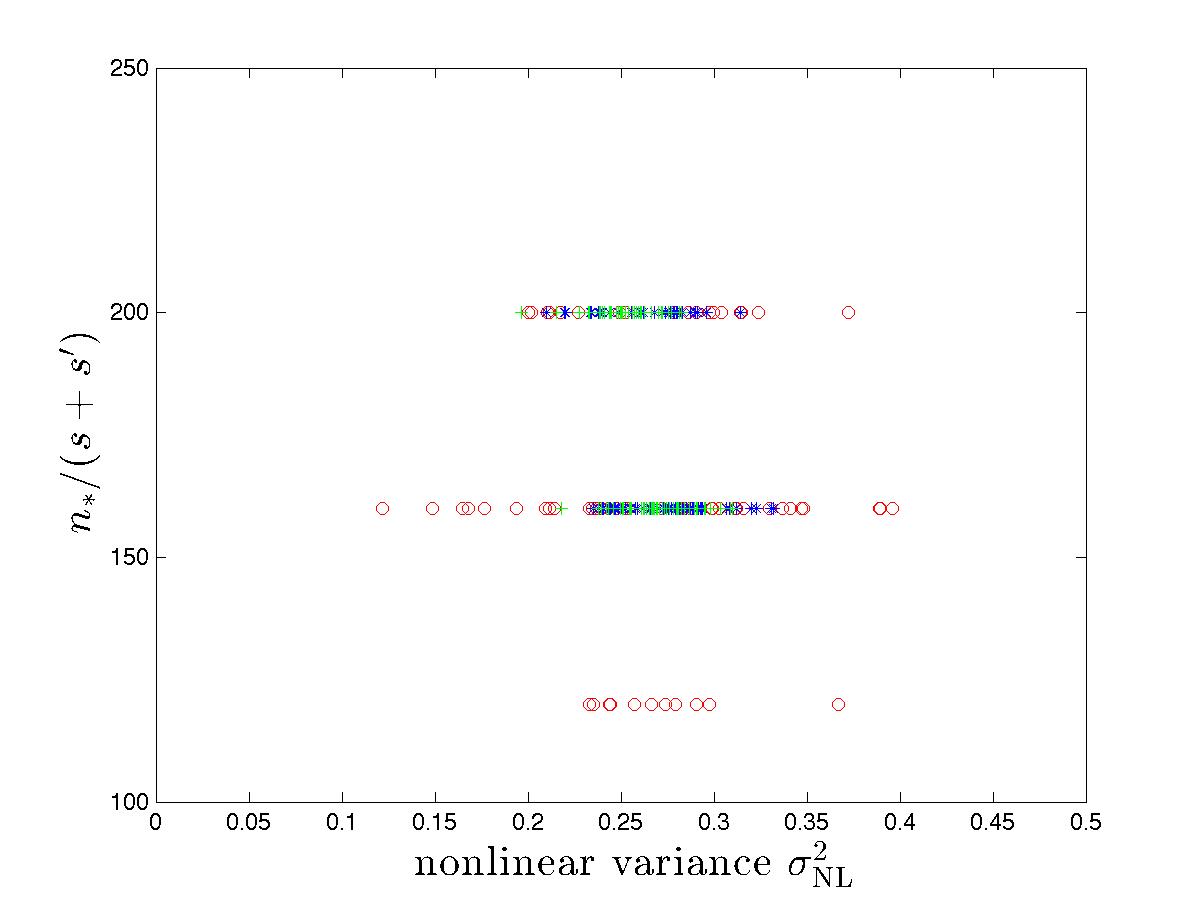}
\caption{PS model with synthetic genomes. Red, blue and green symbols correspond to cases with $s+s'=5, 50, 100$ respectively. Results for 100 runs (i.e., 100 different realizations of the model) are shown for each case.}
\label{fig:PS}
\end{figure}

In Step 1, we scan across increasing sample size $n$ and compute the $p$-values of all genetic markers that have nonzero support (i.e., for which LASSO returns a nonzero value in $\hat{x}$) in order to detect the phase transition in CS performance. The process is terminated when the median $p$-value and the absolute value of its first derivative are both $10^6$ times smaller than the corresponding quantities when the scanning process first starts. (The choice of $10^6$ is arbitrary but worked well in our simulations -- the purpose is merely to detect the region of sample size where the algorithm is working well.) This terminal sample size is defined to be $n_\ast$. The typical behavior of median $p$-value against $n$ is illustrated in Fig. (\ref{fig:Pvalue}). The median $p$-value undergoes a phase transition, dropping to small values. $n_\ast$ as defined above is typically about (2-3) times as large as the sample size at which this first occurs.

In Table (\ref{table}) we display the distribution of false positives found by Step 1. That is, loci to which the algorithm assigns a nonzero effect size when in fact the original model has effect size zero. When false positives are present they cause the subspace explored in Step 2 of the algorithm to have higher dimension than necessary. However, they do not necessarily lead to actual false positives in the final nonlinear model produced by Step 2. As is clear from Table (\ref{table}),
in the large majority of cases the number of false positives is small compared to the number of true positives.

For the BD and PS models, we calculate $n_\ast/s$ and $n_\ast/(s+s')$ which are plotted against $\sigma^2_{\textrm{NL}}$ in Fig. (\ref{fig:BD}) and Fig. (\ref{fig:PS}). Next, we run Step 2 over the causal subspace determined by Step 1. Running CS gives us $\hat{X}$ (as defined in \eqref{bigX})  and then we compute the residual variance:
\bea
\sigma^2_{\textrm R} \equiv \textrm{var} (y - G \,\hat{X} - \epsilon).
\eea
(Recall that $\hat{X}$ means the candidate vector for the solution $X$, as was the case for $\hat{x}$ and $x$.)

In Fig. (\ref{fig:BD}), we display $\sigma^2_R$ against $\sigma^2_{\textrm{NL}}$ and fractions of zeros in both of $x^\ast$ and $\hat{X}$ for the BD model. We also plot $n_\ast/s$ against $\sigma^2_{\textrm{NL}}$. For the PS model, we display the analogous results in Fig. (\ref{fig:PS}).

\begin{figure}[t!]
\includegraphics[height=6.0cm, width=8.0cm]{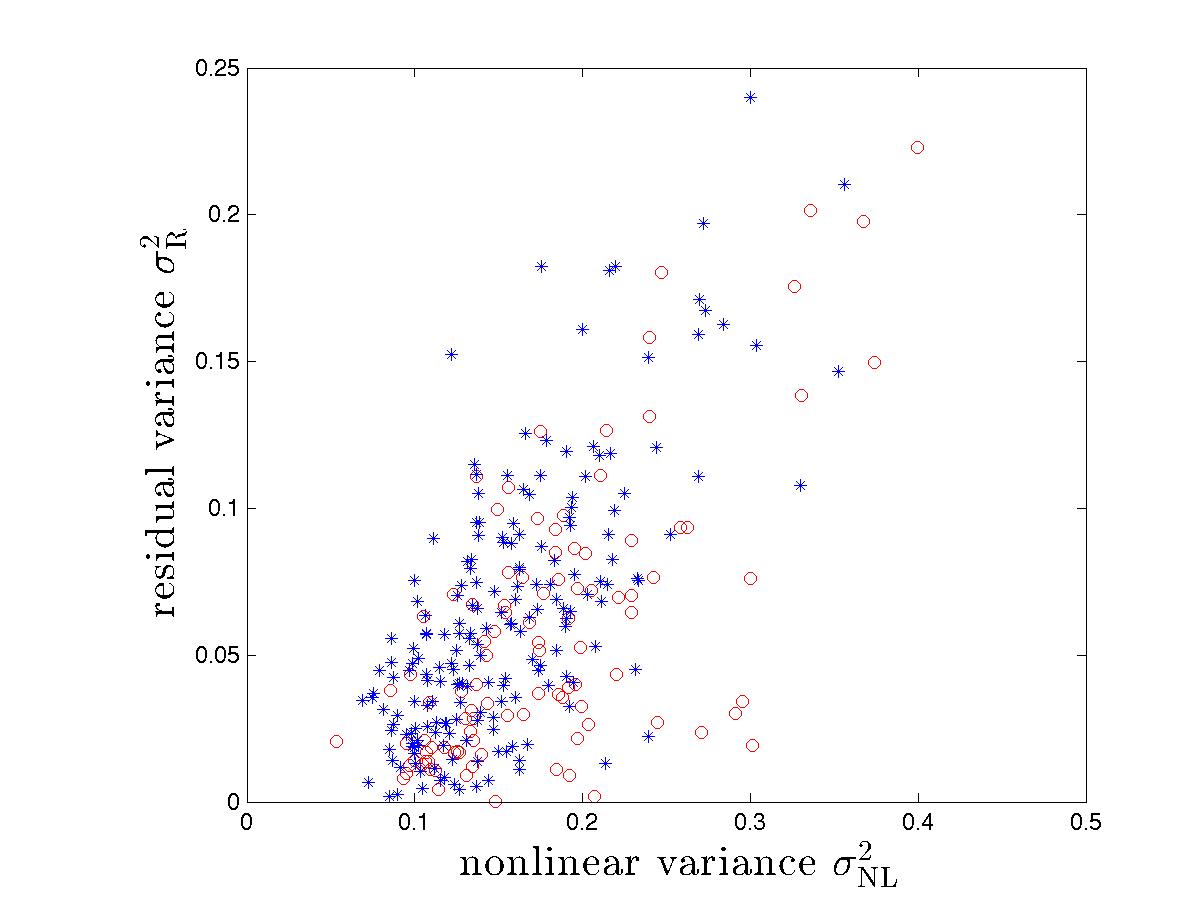}
~~
\includegraphics[height=6.0cm, width=8.0cm]{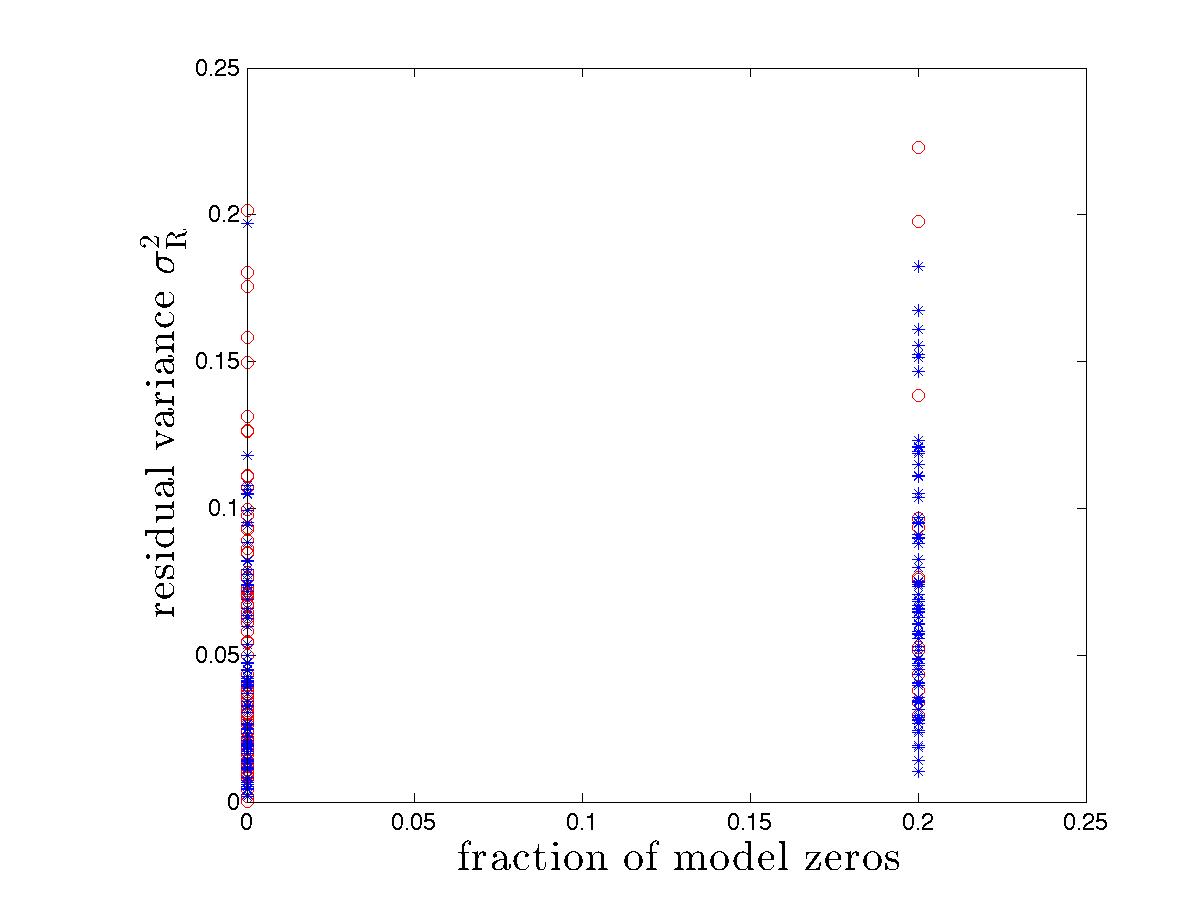}
~~
\includegraphics[height=6.0cm, width=8.0cm]{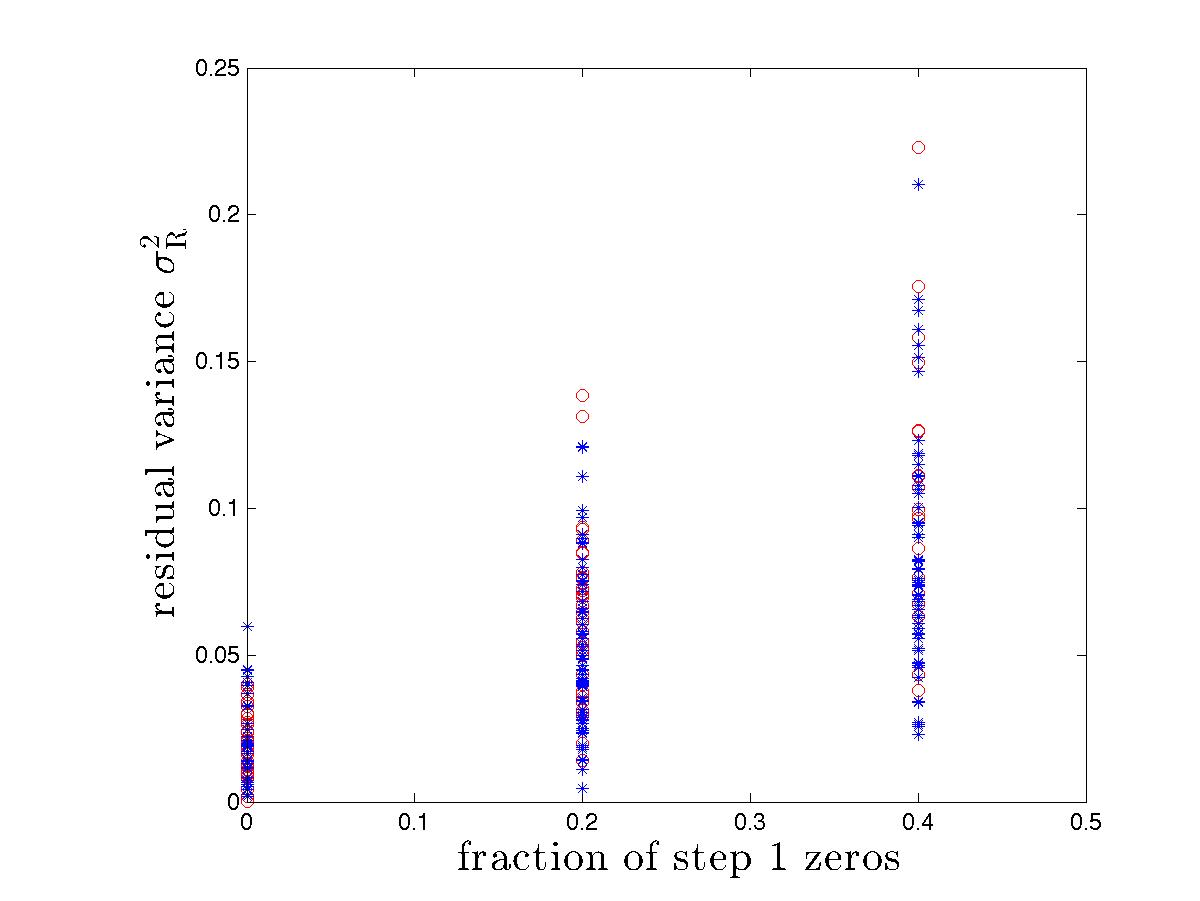}
~~
\includegraphics[height=6.0cm, width=8.0cm]{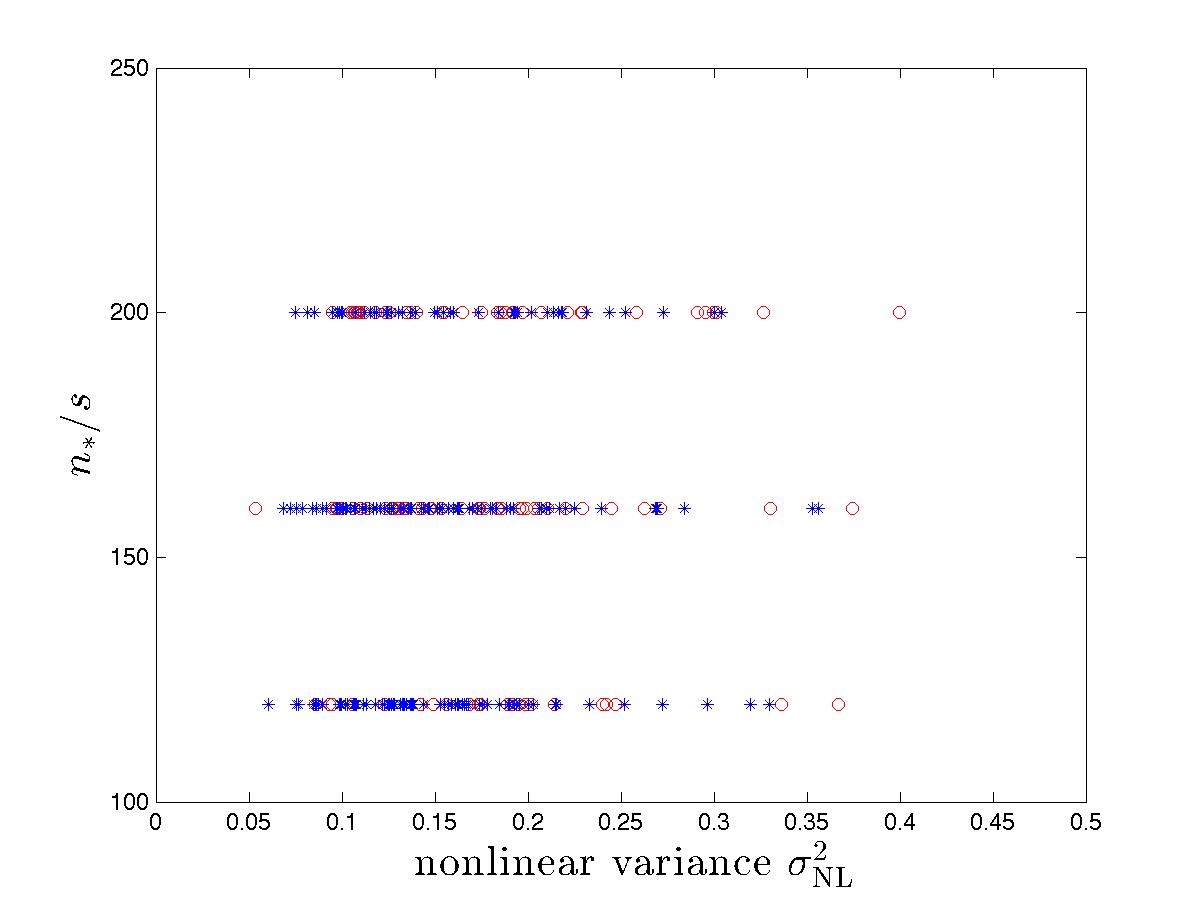}
\caption{Synthetic (red) and real (blue) genome results in the BD model for $s=5$. Results for 100 runs (i.e., 100 different realizations of the model) are shown.}
\label{fig:DataBD}
\end{figure}

\begin{figure}[t!]
\includegraphics[height=6.0cm, width=8.0cm]{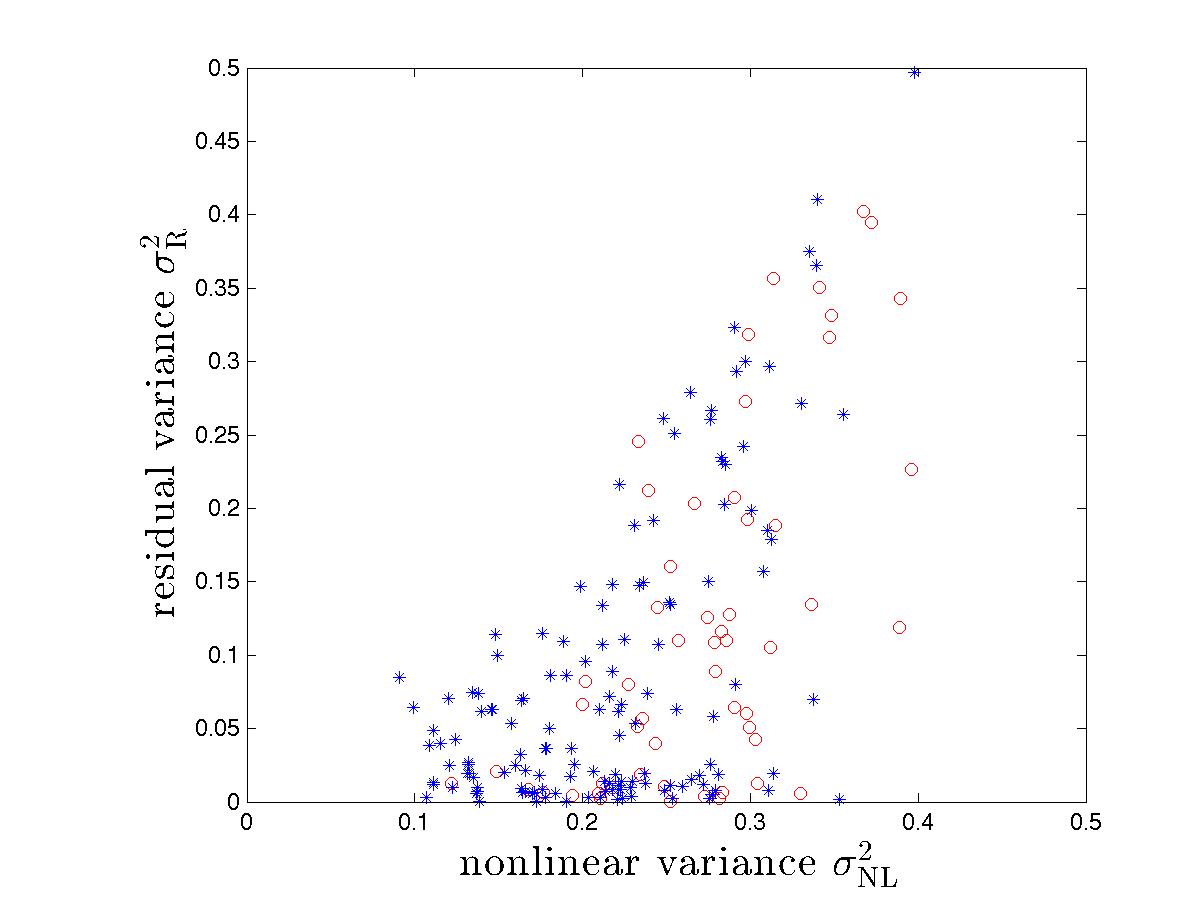}
~~
\includegraphics[height=6.0cm, width=8.0cm]{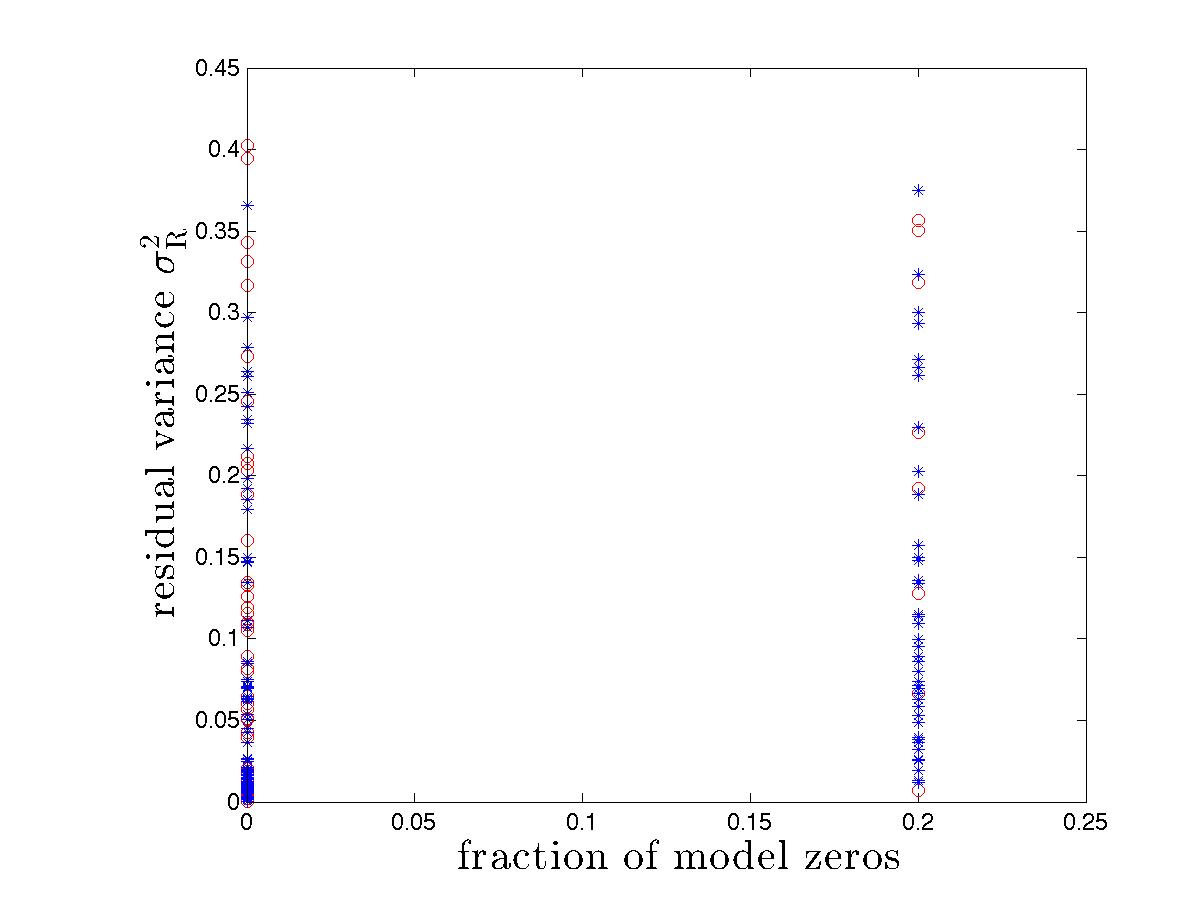}
~~
\includegraphics[height=6.0cm, width=8.0cm]{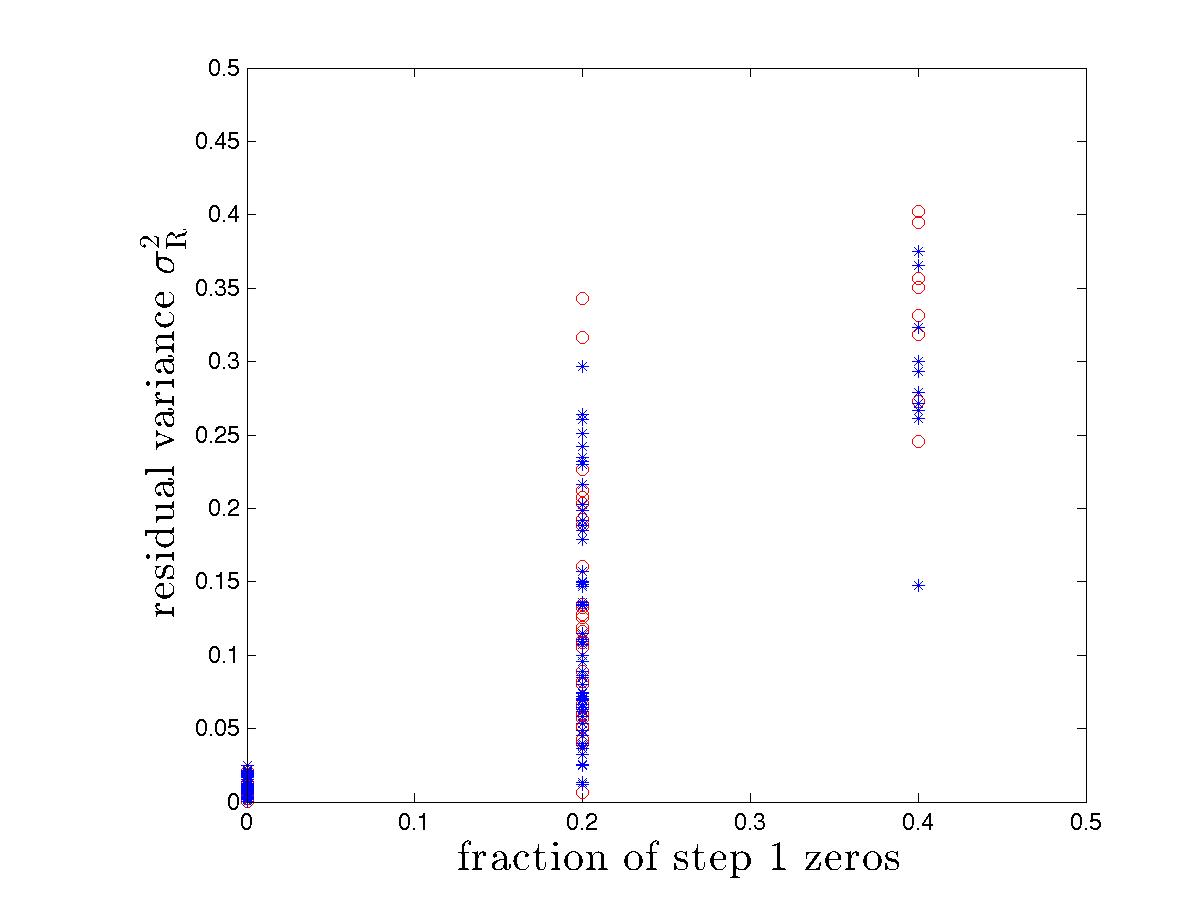}
~~
\includegraphics[height=6.0cm, width=8.0cm]{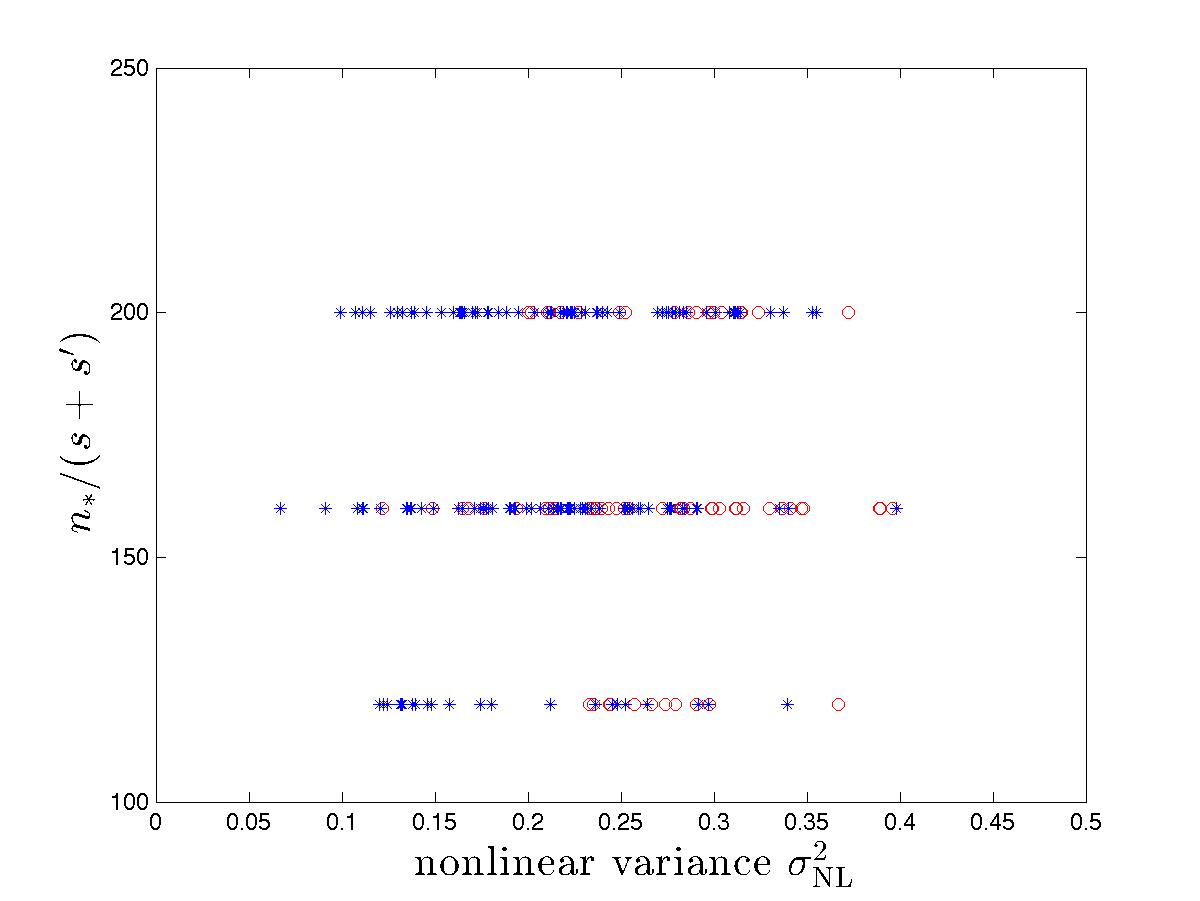}
\caption{Synthetic (red) and real (blue) genome results in the PS model for $s+s'=5$. Results for 100 runs (i.e., 100 different realizations of the model) are shown.}
\label{fig:DataPS}
\end{figure}

Finally, we note that our method performs similarly on synthetic genomes as well as actual human SNP genomes (i.e., matrices $g$ obtained via variant calls on actual genomes from the 1000 Genomes Project). Details
concerning sequencing and SNP calling for the 1000 Genomes Project can be found at: http://www.1000genomes.org/analysis$\,$. For our purposes, we randomly selected p SNPs from each individual to form the rows of our sensing matrix.
In Fig. (\ref{fig:DataBD}) and Fig. (\ref{fig:DataPS}), we compare results on synthetic and real genomes for both BD and PS models. Due to the limited sample size of $\sim 1000$ real genomes to which we have access, we limit ourselves to the cases $s=5$ and $\{s=3,s'=2\}$. In our analysis we could detect no qualitative difference in performance between real and synthetic $g$. However, the correlation between columns of real $g$ matrices is larger than for synthetic (purely random) matrices. To compensate for this, we ran our simulations with slightly larger (e.g., 1.5 or 2 times larger) penalization $\lambda$ in the case of real genomes. Our results suggest, as one might expect from theory, that the moderate correlations between SNPs found in real genomes do not alter the universality class of the compressed sensor $g$.

\section{Method}

Consider the most general model which includes gene-gene interactions (we include explicit indices for clarity; $1 \leq a \leq n$ labels individuals and $1 \leq i,j \leq p$ label genomic loci)
\begin{equation}
\label{model}
y^a = \sum_i g^a_i \,z_i ~+~  \sum_{ij} g^a_i \,Z_{ij} \,g^a_j ~+~ \epsilon^a ~~~,
\end{equation}
where $g$ is an $n \times p$ dimensional matrix of genomes, $z$ is a vector of linear effects, $Z$ is a matrix of nonlinear interactions, and $\epsilon$ is a random error term. We could include higher order (i.e., gene-gene-gene) interactions if desired.

Suppose that we apply conventional CS to data generated from the model above. This is equivalent to finding the best-fit linear approximation
\begin{equation}
\label{yax}
y^a \approx \sum_i g^a_i \,{x}_i~~.
\end{equation}
If enough data (roughly speaking, $n >> s \log p$, where $s$ is the sparsity of $x$) is available, the procedure will produce the best-fit hyperplane approximating the original data.

It seems plausible that the support of $x$, i.e., the subspace defined by nonzero components of $x$, will coincide with the subset of loci which have nonzero effect in {\it either} $z$ {\it or} $Z$ of the original model. That is, if the phenotype is affected by a change in a particular locus in the original model (either through a linear effect $z$ or through a nonlinear interaction in $Z$), then CS will assign a nonzero effect to that locus in the best-fit linear model (i.e., in $x$). As we noted in the Results section, this hypothesis is largely correct: the support of $x$ tends to coincide with the support of $(z,Z)$ except in some special cases where nonlinearity masks the role of a particular locus.

Is it possible to do better than the best-fit linear effects vector $x$? How hard is it to reconstruct both $z$ and $Z$ of the original nonlinear model? This is an interesting problem both for genomics (in which, even if the additive variance dominates, there is likely to be residual non-additive variance) and other nonlinear physical systems.

It is worth noting that although (\ref{model}) is a nonlinear function of $g$ -- i.e., it allows for epistasis, gene-gene interactions, etc. -- the phenotype $y$ is nevertheless a {\it linear} function of the parameters $z$ and $Z$. One could in fact re-express (\ref{model}) as
\begin{equation}
\label{bigX}
y^a = \sum_i G^a_i (g) \,X_i + \epsilon^a
\end{equation}
where $X$ is a vector of effects (to be extracted) and $G$ the most general nonlinear function of $g$ over the $s$-dimensional subspace selected by the first application of CS resulting in (\ref{yax}). Working at, e.g., order $g^2$, $X$ would have dimensionality $s(s-1)/2 + 2s$, enough to describe all possible linear and quadratic terms in (\ref{model}).

Given the random nature of $g$, it is very likely that $G$ will also be a well-conditioned CS matrix (we verify empirically that that this is the case). Potentially, the number of nonzero components of $X$ could be $\sim s^k$ at order $g^k$. However, if the matrix $Z$ has a sparse or block-diagonal structure (i.e., individual loci only interact with some limited number of other genes, not all $s$ loci of nonzero effect; this seems more likely than the most general possible $Z$), then the sparsity of $X$ is of order a constant $k$ times $s$. Thus, extracting the full nonlinear model is only somewhat more difficult than the $Z = 0$ case. Indeed, the data threshold necessary to extract $X$ scales as $\sim k s \log (s(s-1)/2 + 2s)$, which is less than $s \log p$ as long as $k \log (s(s-1)/2 + 2s) < \log p$.
\bigskip

The process for extracting $X$, which is equivalent to fitting the full nonlinear model in (\ref{model}), is as follows:

{\bf Step 1}. Run CS on $(y,g)$ data, using linear model (\ref{yax}). Determine support of $x$: subset defined by $s$ loci of nonzero effect.

{\bf Step 2}. Compute $G(g)$ over this subspace. Run CS on $y = G(g) \cdot X$ model to extract nonzero components of $X$. These can be translated back into the linear and nonlinear effects of the original model (i.e., nonzero components of $z$ and $Z$).
\bigskip

In the Results section we show that in many cases Steps 1 and 2 lead to very good reconstruction of the original model (\ref{model}) given enough data $n$. A number of related issues are discussed:

a. When can nonlinear effects hide causal loci from linear regression (Step 1)? In cases of this sort the locus in question would not be discovered by GWAS using linear methods.

b. Both matrices $g$ and $G(g)$ seem to be well-conditioned CS matrices. The expected phase transitions in algorithm performance are observed for both Steps.

c. For a given partition of variance between linear (L), nonlinear (NL) and IID error $\epsilon$, how much data $n_{\star}$ is required before complete selection of causal variants occurs (i.e., crossing of the phase boundary for algorithm performance)? Typically if Step 1 is successful then with the same amount of data Step 2 will also succeed.

\section{LASSO optimization}

The L1 penalization (e.g., LASSO) involves minimization of an objective function $O$ over candidate vectors $\hat{x}$:
\begin{equation}
\label{O1}
O = \vert \vert y - A \hat{x} \vert \vert_{L2} + \lambda \vert \vert \hat{x} \vert \vert_{L1}~~,
\end{equation}
where $\lambda$ is the penalization parameter. Since $O$ is convex, a local minimum is also a global minimum. The minimization is performed using
pathwise coordinate descent \cite{Pathwise1,Pathwise2} --- optimizing one parameter (coordinate), $\hat{x}_j$, at a time. This results in a modest computational complexity for the algorithm as a whole.

The solution of each sub-sequence involves a shrinkage operator $\mathcal{S}$ \cite{GCS}:
\bea
\mathcal{S}(\hat{x}_j, \, \lambda) = \left\{
                                       \begin{array}{ll}
                                         \hat{x}_j -\lambda, & \hbox{~~~~if \,$\hat{x}_j > 0$ \,and\, $\hat{x}_j > \lambda$~~;} \\
                                         \hat{x}_j +\lambda, & \hbox{~~~~if\, $\hat{x}_j < 0$ \,and\, $|\hat{x}_j| > \lambda$~~;} \\
                                         0, & \hbox{~~~~if\, $|\hat{x}_j| \leq \lambda$~~,}
                                       \end{array}
                                     \right.
\eea
where $j=1,2, \ldots, p$. The penalization parameter $\lambda$ is estimated according to the following two part procedure. We first choose $\lambda$ as the noise variance. Then after running CS, we obtain the nonlinear variance defined in Eq. \eqref{NL_variance}. The ultimate $\lambda$ used in the simulations is taken to be the sum of the noise variance and the nonlinear variance. In a realistic setting one could set $\lambda$ using an estimate of the additive heritability of the phenotype in question (e.g., obtained via twin or adoption study, or GCTA \cite{GCTA1,GCTA2}). A smaller penalization might be sufficient, and allow the phase boundary to be reached with somewhat less data. We assume convergence of the algorithm if the fractional change in $O$ between two consecutive sub-sequences is less than $10^{-4}$. Note that in the case of real genomes we found that slightly increasing $\lambda$ beyond the values described above yielded somewhat better results.

\section{Discussion}

It is a common belief in genomics that nonlinear interactions (epistasis) in complex traits make the task of reconstructing genetic models extremely difficult, if not impossible. In fact, it is often suggested that overcoming nonlinearity will require much larger data sets and significantly more computing power. Our results show that in broad classes of plausibly realistic models, this is not the case.

We find that the proposed method can recover a significant fraction of the predictive power (equivalently, variance) associated with nonlinear effects. The upper left panels of Fig. (\ref{fig:BD}) and Fig. (\ref{fig:PS}) show that we typically recover half or more of the nonlinear variance. To take a specific example, for $\sigma^2_\textrm{NL} \sim 0.25$ over a third of the total genetic variance $$h^2_\textrm{broad sense} \equiv H^2 = 1 - \textrm{var}( \epsilon ) = 0.7$$ is due to nonlinear effects. (Note total variance of $y$ is defined to be unity.) Step 2 of our method recovers all but $\sigma^2_\textrm{R} \sim 0.1$ of the total genetic variance, using the same amount of data as in the linear Step 1. The fraction of variance which is not recovered by our method is largely due to the causal variants that are not detected by Step 1 of the algorithm -- i.e., the fraction of zeros. These variants would also escape detection by linear regression or essentially any other linear method using the same amount of sample data. We have also calculated false positive rates for Step 1 in our simulations and the results are shown in Table 1. Typically, the number of false positives is much smaller than the number of true positives. In other words, the dimensionality of the subspace explored in Step 2 is nearly optimal in most cases. We are unaware of any method that can recover more of the predictive power than ours using similar sample size and computational resources.

Performance would be even better if sample sizes larger than $n_{\ast}$ (which was somewhat arbitrarily defined) were available. Typically, $n_{\ast} \sim 100 \, \times$ sparsity, where sparsity $s$ is the number of loci identified by Step 1 (i.e., dimensionality of the identified causal subspace). A lower value of $n_{\ast} / s$ might be sufficient if we were to tune the penalization parameter $\lambda$ more carefully. In a realistic setting, one can continue to improve the best fit nonlinear model as more data becomes available, eventually recovering almost all of the genetic variance.

\begin{figure}[t!]
\includegraphics[height=6.0cm, width=8.0cm]{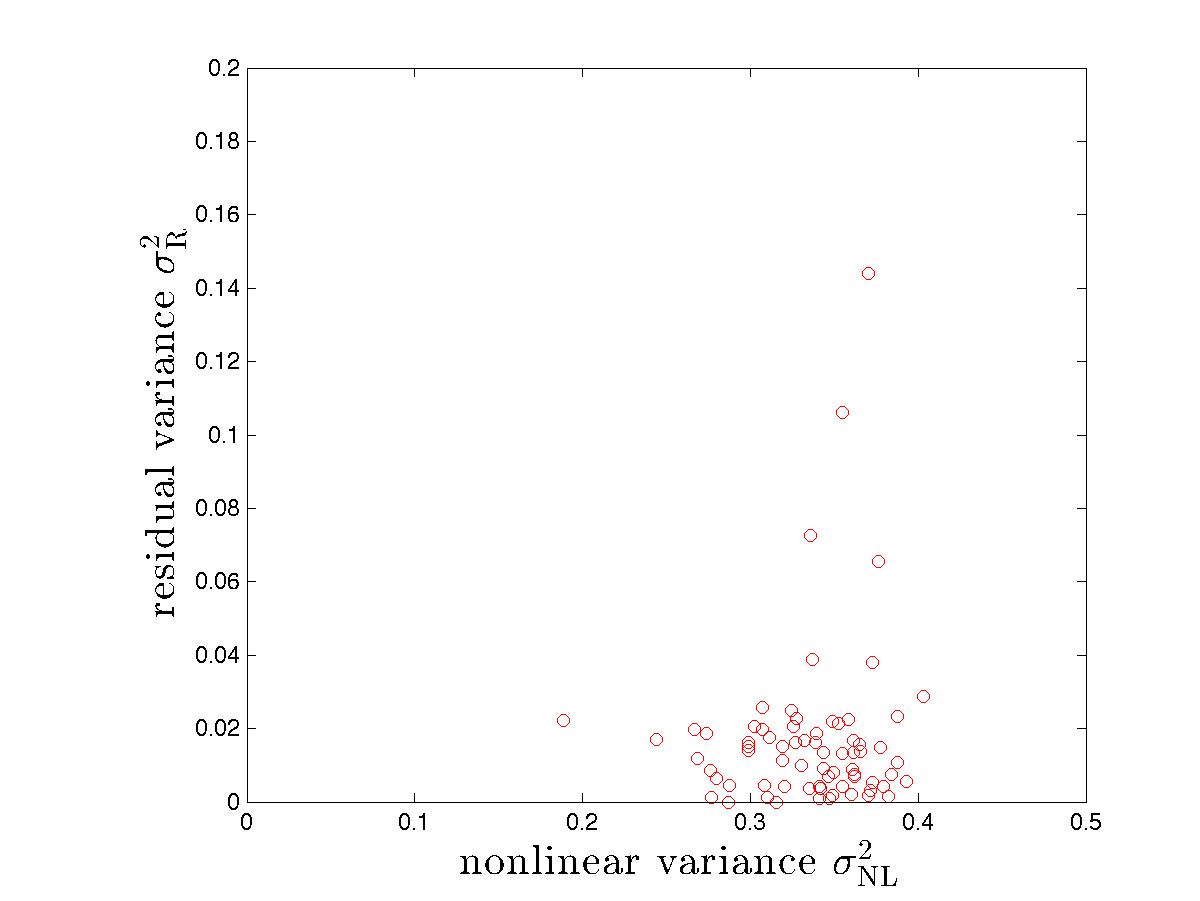}
~~
\includegraphics[height=6.0cm, width=8.0cm]{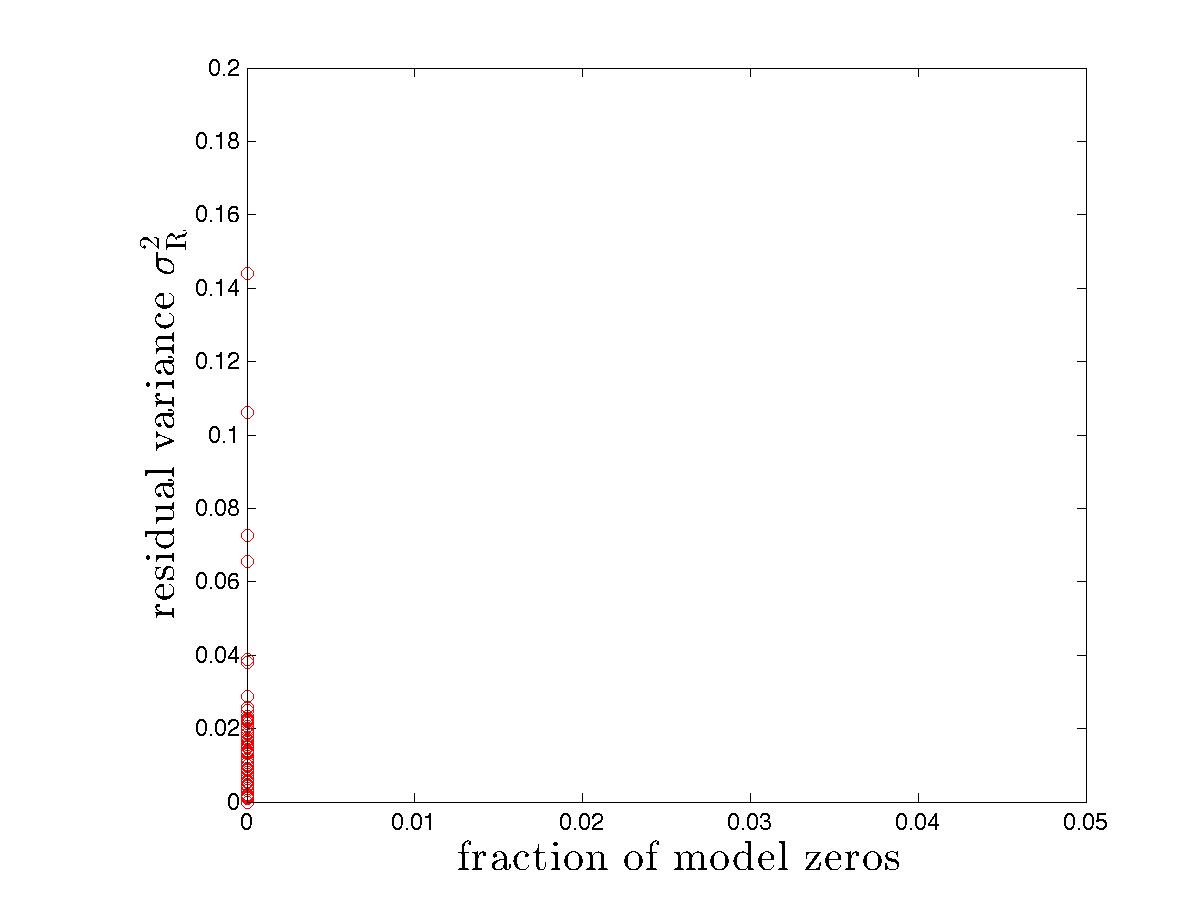}
~~
\includegraphics[height=6.0cm, width=8.0cm]{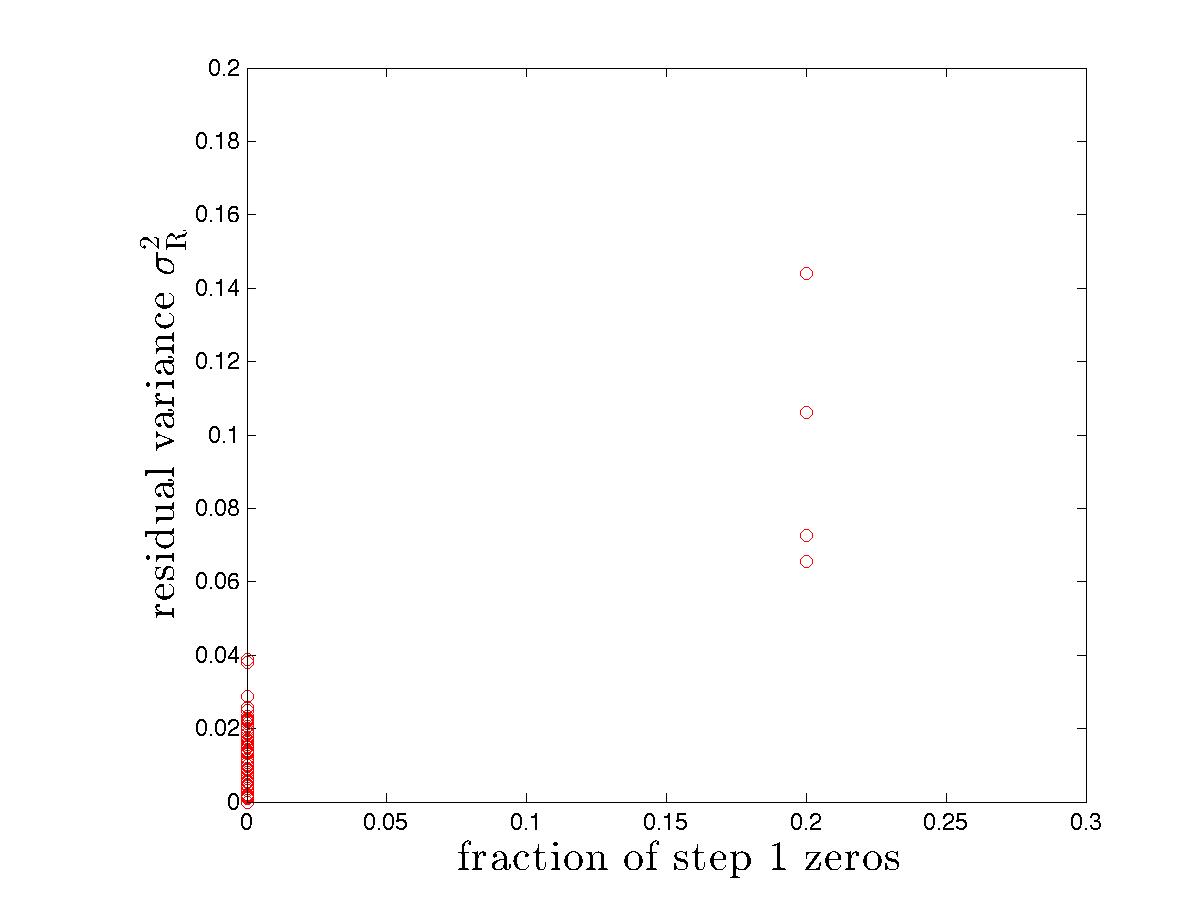}
~~
\includegraphics[height=6.0cm, width=8.0cm]{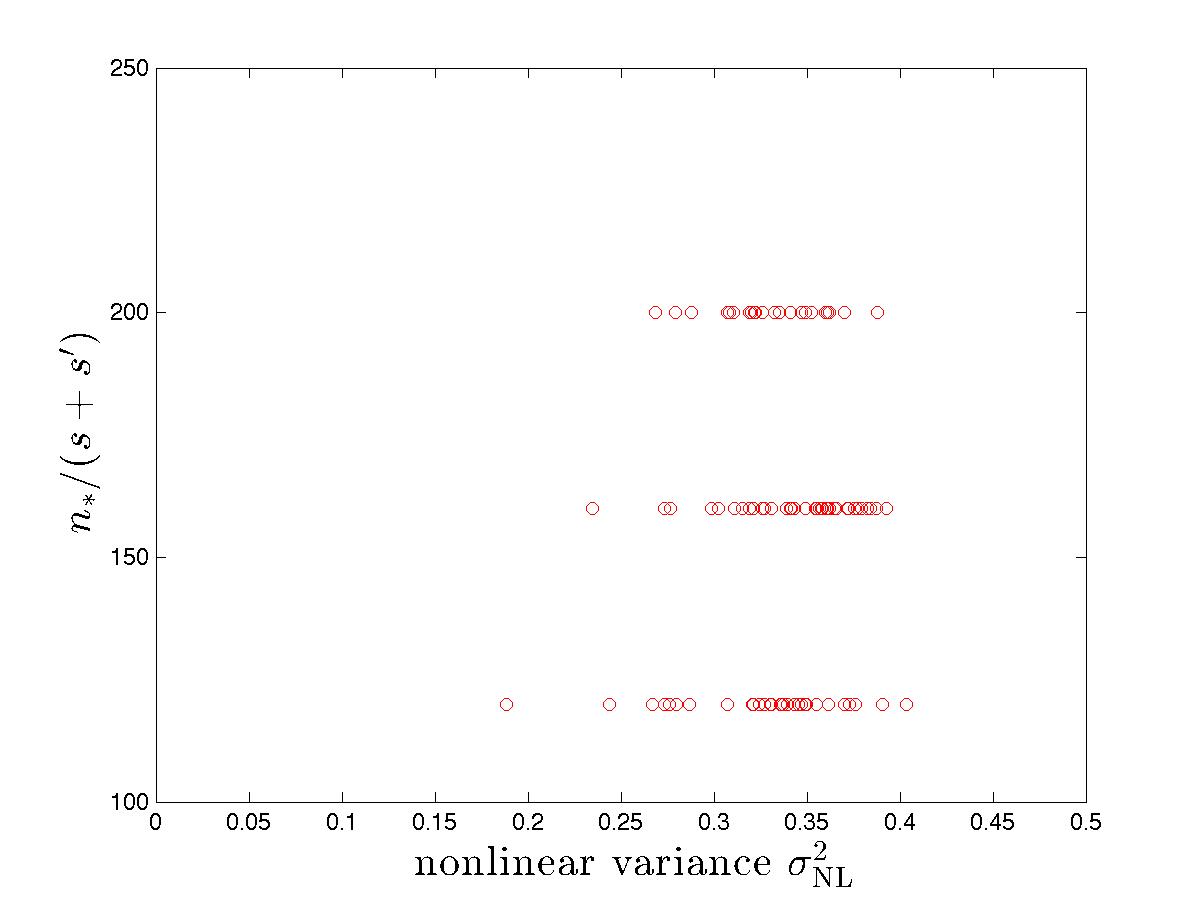}
\caption{The PS model for $s+s'=5$ with continuous $g$ elements. Results for 100 runs (i.e., 100 different realizations of the model) are shown.}
\label{fig:Con}
\end{figure}

Finally, we note that our method can be applied to problems in which the entries in $g$ are continuous rather than discrete.
(For example, compressed sensing is often applied to image reconstruction from scattered light intensities. These intensities are continuously valued and not discrete. It seems possible that the scattering medium might introduce nonlinearities, thereby making our methods of interest.) In fact, discrete values make model zeros more likely than in the continuous case. In Fig. (\ref{fig:Con}), we display the results for a PS model with $\{s=3,s'=2\}$ and matrix entries generated from continuous probability distributions. Recovery of nonlinear variance is generally better than in the discrete case, and the fraction of zeros is smaller.

\section*{Competing interests}
  The authors declare that they have no competing interests.

\section*{Author contributions}
SH conceived the method. CMH performed the numerical simulations. SH and CMH wrote the manuscript.

\section*{Acknowledgements}

%\acknowledgments

The authors thank Shashaank Vatikutti for MATLAB code and help with related issues, and Christopher Chang for SNP genomes based on 1000 Genomes data. The authors also thank Carson Chow, James Lee, and Laurent Tellier for discussions at an early stage of this research. This work is supported in part by funds from the Office of the Vice-President for Research and Graduate Studies at Michigan State University.

% The bibliography will probably be heavily edited during typesetting.
% We'll parse it and, using the arxiv number or the journal data, will
% query inspire, trying to verify the data (this will probalby spot
% eventual typos) and retrive the document DOI and eventual errata.
% We however suggest to always provide author, title and journal data:
% in short all the informations that clearly identify a document.
{}

\newpage

\section{Figures}

\begin{table}[t!]
  \centering
  \begin{tabular}{l*{5}{c}r}
                                                   ~ & FP = 0 ~&  FP = 1  ~& FP = 2  ~& FP = 3 ~& 4 < FP < 7     \\ \hline \hline
BD, synthetic, $s = 5$                             ~ & 0.38   ~& 0.43     ~& 0.17    ~& 0.02   ~& 0     \\ \hline
BD, synthetic, $s = 50$                            ~ & 0.39   ~& 0.37     ~& 0.19    ~& 0.03   ~& 0.02   \\ \hline
BD, synthetic, $s = 100$                           ~ & 0.35   ~& 0.40     ~& 0.18    ~& 0.06   ~& 0.01   \\ \hline
PS, synthetic, $s = 3,\, s' = 2$                   ~ & 0.38   ~& 0.44     ~& 0.14    ~& 0.04   ~& 0     \\ \hline
PS, synthetic, $s = 30,\, s' = 20$                 ~ & 0.24   ~& 0.36     ~& 0.27    ~& 0.09   ~& 0.04      \\ \hline
PS, synthetic, $s = 60,\, s' = 40$                 ~ & 0.42   ~& 0.29     ~& 0.19    ~& 0.06   ~& 0.04       \\ \hline
BD, real data, $s = 5$                             ~ & 0.74   ~& 0.18     ~& 0.05    ~& 0      ~& 0.03   \\ \hline
PS, real data, $s = 3,\, s' = 2$                   ~ & 0.69   ~& 0.13     ~& 0.08    ~& 0.04   ~& 0.06   \\ \hline
Continuous, PS, synthetic, $s = 3,\, s' =2$        ~ & 0.29   ~& 0.35     ~& 0.23    ~& 0.09   ~& 0.04   \\ \hline
\end{tabular}
\caption{Distribution of number of false positives in our simulations. These are loci which Step 1 incorrectly identifies as affecting the phenotype. The first column gives the probability that no false positives are found, the second column gives the probability that one false positive is found, etc. }
\label{table}
\end{table}

\end{document}